\documentclass[fleqn,usenatbib]{mnras}

\usepackage{newtxtext,newtxmath}
\usepackage[T1]{fontenc}
\DeclareRobustCommand{\VAN}[3]{#2}
\let\VANthebibliography\thebibliography
\def\thebibliography{\DeclareRobustCommand{\VAN}[3]{##3}\VANthebibliography}
\usepackage{graphicx}	
\usepackage{amsmath}	
\usepackage{latexsym}
\usepackage{physics}
\usepackage{subfigure}

\title[CAFE-AMR]{CAFE-AMR: A computational MHD Solar Physics simulation tool that uses AMR}

\author[Ricardo Ochoa-Armenta \& Francisco S. Guzm\'an]{
Ricardo Ochoa-Armenta\thanks{E-mail: ricardo.ochoa@umich.mx}
and Francisco S. Guzm\'an\thanks{E-mail: francisco.s.guzman@umich.mx}
\\
Instituto de F\'{\i}sica y Matem\'{a}ticas, Universidad
              Michoacana de San Nicol\'as de Hidalgo. Edificio C-3, Cd.
              Universitaria, 58040 Morelia, Michoac\'{a}n,
              M\'{e}xico.
}

\date{Accepted XXX. Received YYY; in original form ZZZ}

\pubyear{20XX}

\begin{document}
\label{firstpage}
\pagerange{\pageref{firstpage}--\pageref{lastpage}}
\maketitle

\begin{abstract}
The study of our Sun holds significant importance in Space Weather research, encompassing a diverse range of phenomena characterized by distinct temporal and spatial scales. To address these complexities, we developed CAFE-AMR, an implementation of an Adaptive Mesh Refinement (AMR) strategy coupled with a Magnetohydrodynamics (MHD) equation solver, aiming to tackle Solar Physics-related problems. CAFE-AMR employs standard fluid dynamics methods, including Finite Volume discretization, HLL and Roe class flux formulas, linear order reconstructors, second-order Runge-Kutta, and Corner Transport Upwind time stepping. In this paper, we present the core structure of CAFE-AMR, discuss and evaluate mesh refinement criteria strategies, and conduct various tests, including simulations of idealized Solar Wind models, relevant for Space Weather applications.
\end{abstract}

\begin{keywords}
methods: numerical -- MHD -- Sun: general
\end{keywords}


\section{Introduction.}
\label{sec:Introduction}

Solar physics has been developed in a constant direction: much of our Sun's present observed structure and dynamic behavior owe their existence to the magnetic field. In every shell that constitutes the solar atmosphere, being either the photosphere, the chromosphere, the corona, or the heliosphere, there are many endearing and involving questions relating the Sun's plasma to the magnetic field.

\textit{Magnetohydrodynamics} (MHD) describes fairly well several of these magnetic field dynamics features. Fortunately both, analytical and numerical advances in our understanding and use of these equations have been made. Moreover, advances regarding the improvement of the usage of computational resources while running simulations based on these equations have been equally important.

A significant breakthrough on Fluid Dynamics simulations, was made a while ago  by the implementation of \textit{Adaptive Mesh Refinement} (AMR) technique, presented in the well known \citep{berger1984amr} and \citep{berger1989local}, where it was used in conjunction with the numerical shock capturing solutions of the Euler hydrodynamic equations. AMR is an approach to optimize the efficient use of resources in the solution of initial value problems involving the solution of fluid dynamics; it can be summarized as the use of more resolution, spatial or temporal, only where more structure of the functions involved in the problem demand more accuracy. Physically, AMR is a powerful tool to adequately evolve a fluid when great disparities in the spatial and temporal scales occur simultaneously, when the flow dynamics have very localized features.

A number of codes that employ the AMR strategy for solving MHD equations are available within the scientific community. Some examples are the following:
MPI-AMRVAC \citep{porth2014mpi}: This code is specifically designed to solve MHD and Hall MHD equations, it utilizes finite volume or finite difference methods and can also handle the evolution of dust coupled with a hydrodynamic fluid. A second \citep{xia2018mpi} and third \citep{keppens2023mpi} versions of this code have been developed; PLUTO \citep{mignone2007pluto}: which has been used mainly in astrophysical simulations and has gained attraction in the field of solar physics in recent years; FLASH\citep{fryxell2000flash}: A widely used astrophysical code; AstroBEAR \citep{cunningham2009simulating}; RAMSES \citep{teyssier2002cosmological};  NIRVANA \citep{nirvana}; ATHENA\citep{stone2008athena}; ENZO \citep{bryan2014enzo} and BATS-R-US \citep{toth2012adaptive} are codes known for their high-performance computing orientation and have been extensively used in various scientific disciplines, including astrophysics and MHD simulations. Some of these codes rely on specialized libraries such as PARAMESH \citep{macneice2000paramesh} and CHOMBO \citep{CHOMBO}, which provide parallelized AMR drivers that can be adapted to solve initial value problems as needed. In the realm of solar physics simulation, noteworthy codes include AMR-CESE-MHD \citep{feng2012validation}, ICARUS \citep{verbeke2022icarus}, which support non cartesian grids, and SFUMATO \citep{matsumoto2019dynamical}, all of which have been successfully used in solar wind simulations.

The AMR strategy, when integrated with parallelization schemes, significantly enhances computational efficiency. State-of-the-art computing systems often adopt a combination of technologies. For example OpenMP-MPI, as demonstrated in the BHAC GR-MHD code \citep{cielo2022optimizing}. Also, the growing adoption of GPUs is expected to further amplify the processing power for handling vast data sets generated by computer simulations. A notable example of this trend is the H-AMR code \citep{liska2022h}.

Even though most of these codes are open source, the complexity of their modular structure might render them inaccessible. As an alternative, developing an original code based on standard numerical schemes, allows one to have an independent tool that serves as a theoretical laboratory that on the one hand helps to confirm results obtained with other codes, and on the other serves to implement the solution of new problems that would be hard to define on top of  already existing codes that, due to their degree of sophistication prevent easy soft adjustments. Accessibility and control are an important motivation to build a code.

In this work we present a code, constructed from scratch, that solves the equations of the MHD using AMR based on the methods for the solution of the equations developed for the unigrid original version of CAFE \citep{CAFE,CAFEN}. This code is expected to 
join those others that are in production for Solar and Space Weather Physics. The paper structure is as follows. In section \ref{sec:MHD eq and Numerical methods}, an overview of resistive MHD equations and the numerical methods is presented. Section \ref{sec:AMR technique} is an exposition of the adaptive mesh refinement strategy. In Section \ref{sec: Code Structure} we describe the numerical methods we use for the solution of the MHD equations and the AMR strategy. In Section \ref{sec:tests} we present standard and specialized tests on Solar Physics. Finally in Section \ref{sec:conclusions} we present final comments.

\section{MHD equations and numerical methods.}
\label{sec:MHD eq and Numerical methods}

CAFE-AMR has been designed as a primer to solve time dependent initial value problems described by a set of quasilinear hyperbolic time dependent differential equations. This was in consideration of that resistive MHD equations can be written in a way to satisfy hyperbolicity criteria. 

\subsection{MHD equations.}
\label{subsec: Ideal MHD}

The set of MHD equations we are using to describe the dynamics of a fluid element are the following:

\begin{equation}\label{eq: mass conservation}
\dfrac{\partial \rho}{\partial t}+\nabla\cdot(\rho \mathbf{v})=0,    
\end{equation}
\begin{equation}\label{eq:Momentum conservation BClean}
\dfrac{\partial \rho \mathbf{v}}{\partial t}+\nabla\cdot[\rho \mathbf{v}\mathbf{v}-\mathbf{B}\mathbf{B}]+\nabla p_t=\left(\nabla\cdot\mathbf{B}\right)\mathbf{B} -\rho \nabla \Phi,
\end{equation}
\begin{equation}\label{eq:Energy conservation BClean}
\begin{split}
\dfrac{\partial \varepsilon}{\partial t}+\nabla\cdot[(\varepsilon + p_t)\mathbf{v}-(\mathbf{v}\cdot\mathbf{B})\mathbf{B}]=-\nabla\left(\eta\mathbf{J}\times\mathbf{B}\right)\\
-\rho\mathbf{v}\cdot\nabla\Phi -\mathbf{B}\cdot(\nabla\psi),    
\end{split}
\end{equation}
\begin{equation}\label{eq: ohm Law BClean}
\dfrac{\partial \mathbf{B}}{\partial t}-\nabla\times(\mathbf{v}\times\mathbf{B})=\nabla\times [\eta\mathbf{J}],
\end{equation}
\begin{equation}\label{eq: Psi}
\partial_t\psi+C_h^2\nabla\cdot\mathbf{B}=-\frac{C_h^2}{C_p^2}\psi.
\end{equation}

\noindent where $\rho$ is the plasma mass density, $\mathbf{v}$ is the local fluid velocity, $\mathbf{B}$ is the magnetic field, $\Phi$ is the gravitational potential, $\varepsilon$ is the total energy density, $P_t=(\gamma-1)(\varepsilon +\rho\mathbf{v}^2/2+\mathbf{B}^2/2$) is the total pressure, $\mathbf{J}$ is the electric current density and $\eta$ is the resistivity. 

We assume the thermal pressure $p$ of the plasma is given by the equation of state of an ideal gas so that the total energy density is

\begin{equation}
\varepsilon=\frac{p}{\gamma-1}+\frac{1}{2}\rho\mathbf{v}^2+\frac{1}{2}\mathbf{B}^2,
\end{equation}

\noindent where $\gamma$ is the adiabatic index.

In order to maintain the solenoidal constraint, that is $\nabla\cdot\mathbf{B}=0$, we implement the divergence cleaning scheme proposed in \citep{dedner2002hyperbolic}, where the divergence of $\mathbf{B}$ is related to the scalar function $\psi$ that satisfies equation \eqref{eq: Psi}. With this correction, numerical errors that produce a non zero divergence of the magnetic field are transported and diffused through the numerical domain, and at the same time the addition of the equation for $\psi$ maintains the hyperbolic-parabolic structure of the equations. The coefficients $C_h$ and $C_p$ in Eq. (\ref{eq: Psi}) determine the damping of the divergence of the magnetic field.

\subsection{Numerical methods.}
\label{subsec:numericalmethods}

The system of equations \eqref{eq: mass conservation}, \eqref{eq:Momentum conservation BClean}, \eqref{eq:Energy conservation BClean}, \eqref{eq: ohm Law BClean}, and \eqref{eq: Psi} is solved as an Initial Value Problem (IVP) defined on the domain described in Cartesian coordinates $D\times[0,t_f]:=[x_{min},x_{max}]\times[y_{min},y_{max}] \times[z_{min},z_{max}]\times[0,t_f]$, provided initial and boundary conditions are given. We define a base cell centered discretization of $D$ as follows:

\begin{eqnarray}
x_{i} &=& x_{min}+\left(i-\frac{1}{2}\right)\Delta x\nonumber\\
y_{j} &=& y_{min}+\left(j-\frac{1}{2}\right)\Delta y\nonumber\\
z_{k} &=& z_{min}+\left(k-\frac{1}{2}\right)\Delta z\nonumber
\end{eqnarray}

\noindent where $(x_i,y_j,z_k)$ is the center of a cell of volume $\Delta x \Delta y \Delta z$; the indices take values $i=1,...,N_x$, $j=1,...,N_y$, $k=1,...,N_z$; the spatial resolutions are defined by 
$\Delta x = (x_{max}-x_{min})/N_x$, 
$\Delta y = (y_{max}-y_{min})/N_y$ and 
$\Delta z = (z_{max}-z_{min})/N_z$, with $N_x,N_y,N_z$ are the numbers of cells along each direction. Time is discretized by $t^n = n\Delta t$ where time resolution is $\Delta t=CFL\times \mathrm{min}[\Delta x,\Delta y,\Delta z]$, where $CFL$ is the Courant-Friedrichs-Lewy factor. In this domain the value of a given grid function $f$ at time $t^n$ and position $(x_{i},y_{j},z_{k})$ of the numerical domain, will be represented by $f^n_{i,j,k}$.

For the system of equations to be solved we define the vector of conservative variables ${\bf U}=(\rho,\rho{\bf v},\varepsilon, {\bf B},\psi)^T$. The finite volume implementation uses a Godunov type of shock capturing flux computation scheme, which prescribes the evolution formula for ${\bf U}$ from time $t^n$ to $t^{n+1}$ as:

\begin{equation}\label{Flux}
\begin{split}
    \hat{\mathbf{U}}^{n+1}_{i,j,k}&=\hat{\mathbf{U}}^{n}_{i,j,k}\\
    {}&+\Delta t\left[\frac{1}{\Delta x}\left(\mathbf{F}^n_{i-1/2,j,k}-\mathbf{F}^n_{i+1/2,j,k}\right) \right.\\
    {}&+ \frac{1}{\Delta y}\left(\mathbf{G}^n_{i,j-1/2,k}-\mathbf{G}^n_{i,j+1/2,k}\right)\\
    {}&+ \frac{1}{\Delta z}\left(\mathbf{H}^n_{i,j,k-1/2}-\mathbf{H}^n_{i,j,k+1/2}\right)\\
    {}&+\left.\frac{}{} \mathbf{S}^n_{i,j,k} \right]
\end{split}
\end{equation}

\noindent where $\mathbf{F}$, $\mathbf{G}$, $\mathbf{H}$ are the numerical fluxes across the faces of each cell perpendicular to $ x$, $ y$ and $ z$ directions respectively, as specified by the subindices $i\pm 1/2$, $j\pm 1/2$, $k\pm 1/2$. These fluxes  are calculated using either the HLLE \citep{harten1983upstream} \citep{einfeldt1988godunov} or Roe \citep{roe1981approximate} \citep{powell1999solution} approximate Riemann solvers.

Time evolution is implemented using the method of lines; we have two second order implementations to do so. A TVD second order Runge-Kutta method consistent with cell centering in the space-time volume with an appropriate reconstructor for the inter-cell boundary values; among the reconstructors available in CAFE-AMR, in the tests below we use the linear piecewise second-order Minmod, methods. The other time evolution scheme we have worked out leans on the corner transport Upwind \citep{colella1990multidimensional} that implements the GLM-MHD equations, using the schemes proposed in \citep{mignone2010second}.

Following the guidelines in \citep{dedner2002hyperbolic}, for the hyperbolic cleaning, the divergence present in the source term is implemented directly into the equations with the addition of the exact solution for the normal component of the magnetic field of the intercell face and $\psi$. We use the operator splitting on the parabolic part of the source term, which has an exact solution.

\begin{equation}\label{eq:Parabolic Divergence Cleaning}
\psi^{n+1}_{i,j,k}=\exp\left[{-\frac{c_h^2}{c_p^2}}\Delta t\right]\psi^{*n+1}_{i,j,k},
\end{equation}

\noindent where $\psi^{*n+1}_{i,j,k}$ is the evolution of $\psi^{n}$ only through the hyperbolic part of the equations. As in \citep{dedner2002hyperbolic} we use the parameter $C_r$ to control the parabolic coefficient $C_p$ such that

\begin{equation}\label{eq:C_r}
C_r=\frac{C_h}{C_p}
\end{equation}

\noindent this choice has the overall effect of maintaining a homogeneous decay of the function $\psi$ throughout the whole numerical domain. Throughout the tests in this work, $C_r$ is set to 0.9. The other source terms present in the right hand side of  equations (\ref{eq: mass conservation})-(\ref{eq: Psi}), are calculated every time step and in cases when they contains spatial derivatives, these are computed using the centers of each cell for a point-wise second order finite difference approximation.

\section{AMR}
\label{sec:AMR technique}

The Adaptive Mesh Refinement strategy for solving time dependent partial differential equations on rectangular grids, consist of covering the whole spatial domain with a global \textit{coarse} grid, and grid patches that cover certain regions with smaller cells and in doing so defining a 
\textit{finer} mesh on each patch. With this strategy, the numerical time integrator can be applied indistinctly to all of the patches and also the global coarse grid, so that evolution of every subdomain may be done independently.

The implementation uses the standard Berger-Oliger algorithm 
\citep{berger1984amr} and consists of the following steps:

\begin{enumerate}
    \item Select regions of $D$ to be refined.
    \item Generate refined and properly nested grids in such regions.
    \item  Interpolate the numerical values of the conservative variables at each cell of the refined regions.
    \item Once all of the refined grids have been generated, advance in time the domain $D$ as well as the nested grids independently until all of them are synchronized. 
    \item Inject the values of variables calculated on the refined domain into the cells of the coarse domain.
    \item In order to maintain flux balance, correct the fluxes at the boundaries between refined and coarse regions.   
\end{enumerate}

This sequence may be used recursively and applied to subsequent \textit{refinement levels}, that is, all the refined patches can be further refined; in this paper, we will be using the label $\ell$ as the superscript to denote the refinement levels. The resolution ratio $r$, between consecutive refinement levels $\ell$ and $\ell+1$ is set to $r=2$, so that for every cell of the coarse grid, there will be four cells of the fine grid for a problem defined on two space dimensions, and eight in the case of a problem in three spatial dimensions.

The reminder of this section includes a detailed description of each of these steps.

\subsection{Selecting a region to refine.}
\label{subsec: Selecting a region to refine}

The selection of the sectors within the numerical domain to be refined need a {\it refinement criterion}. There is a certain liberty to define this criterion conveniently, according to the physical scenario that will be simulated. 
Typically, an arbitrary number of positive finite scalar functions $\chi(\mathbf{U}^{\ell})$ are defined in every cell of the coarse grid, then every cell $i,j,k$ is flagged to be refined if this scalar function has a value greater than a certain threshold $\chi_r$:

\begin{equation}\label{Rcriteria}
    \chi_{i,j,k}>\chi_r.
\end{equation}

Among others, in this code we implement two principal refinement criteria:

\begin{itemize}
\item Choosing as the evaluating functions the same numerical approximation of the gradients of every conservative variable as the one used to linearly interpolate their values in a new refined grid. That is:

\begin{equation}\label{RcritMio}
    \chi_{m}=\overline{\nabla}(U_m),
\end{equation}

\noindent where $U_m$ is the $m-th$ conservative variable.

\item The refinement criterion used in \citep{mignone2011pluto}, based on the second derivative error norm defined in \citep{lohner1987adaptive}, where

\begin{equation}\label{ref criteria}
\chi(\mathbf{U})=\sqrt{\frac{\sum\limits_{d\in\{x,y,z\}}|\Delta_{d,+1/2}\sigma-\Delta_{d,-1/2}\sigma|^2}{\sum\limits_{d\in\{x,y,z\}}(|\Delta_{d,+1/2}\sigma|+|\Delta_{d,-1/2}\sigma|+\epsilon\sigma_{d,ref})^2}},
\end{equation}

\noindent and $\sigma \equiv \sigma(\mathbf{U})$ is a function of the conservative variables, $\Delta_{d,\pm 1/2}\sigma$ are the undivided forward and backward finite differences along the direction $d$, for example

\begin{displaymath}
\Delta_{x,\pm 1/2}\sigma=\pm(\sigma_{i\pm 1}-\sigma_i).
\end{displaymath} 

\noindent The last term that appears in the denominator is given by: 

\begin{displaymath}
\sigma_{x,ref}=|\sigma_{i+1}|+2|\sigma_{i}|+|\sigma_{i-1}|,
\end{displaymath}
\begin{displaymath}
\sigma_{y,ref}=|\sigma_{j+1}|+2|\sigma_{j}|+|\sigma_{j-1}|,
\end{displaymath}
\begin{displaymath}
\sigma_{z,ref}=|\sigma_{k+1}|+2|\sigma_{k}|+|\sigma_{k-1}|,
\end{displaymath}

\noindent adding the $\sigma_{d,ref}$ term multiplied by the factor $\epsilon<1$ prevents the refinement of regions where small ripples are found.
\end{itemize}

Additionally, when a simulation uses more refinement levels, in order to properly nest them after refinement level $\ell$, cells are flagged according to the refinement criterion, cells in the level $\ell+1$ that contain cells of level $\ell+2$ are also flagged, and finally, level $\ell$ cells that have flagged neighbors of the same level are flagged as well. This procedure is made recursively from the most refined grid to the coarsest one.

\subsection{Generating a new refined grid}

\begin{figure}
    \centering
    \includegraphics[scale=0.5]{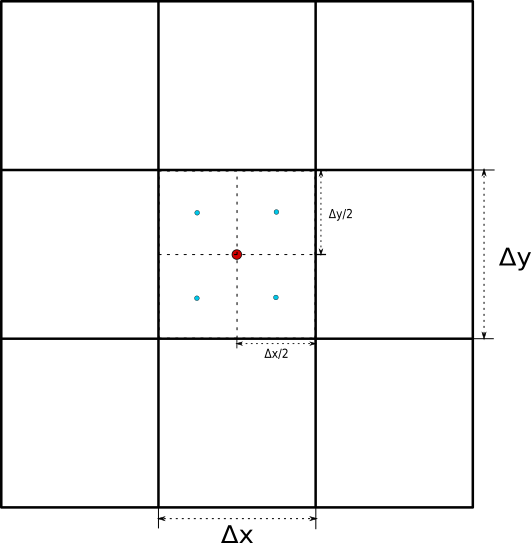}
    \caption{Illustration of how the cell volume of refinement level $\ell+1$, that is $\mathrm{V}^{\ell+1}$ = $\Delta x/2 \times \Delta y/2$, is one quarter of the cell volume in refinement level $\ell$, that is $\mathrm{V}^{\ell}$ = $\Delta x \Delta y$,  for a refinement factor $r=2$ and a dimension of the domain $d=2$.}
    \label{Cell position}
\end{figure}

A mentioned before, the refinement factor we use this work is $r=2$, so that a coarse cell will be covered with $2^d$ cells of the refined domain, where $d$ is the dimension of the domain. In Figure \ref{Cell position} it is shown how the cell volume of refinement level $\ell+1$ is $1/2^d$ of a cell in level $\ell$.

The refined cells will be centered in positions such that the region covered by them has the same boundaries as the coarse cell. For example, in two dimensions, four refined cells that cover a coarse one will have their centers at positions:

\begin{equation}\label{Cell center location}
    \mathbf{r}^{\ell+1}_{I\pm i,J\pm j}=\mathbf{r}^{\ell}_{I,J}\pm\frac{\Delta x}{4}\hat{x}\pm\frac{\Delta y}{4}\hat{y},
\end{equation}

\noindent where $\mathbf{r}^{\ell}_{I,J}$ is the coarse cell's center.

In what follows we use $D^{\ell}$ to denote the domain covered with the grid of refinement level $\ell$.

\subsection{Interpolation of data from coarse to refined grid.}
\label{subsec:Interpolating data}

Likewise in \citep{porth2014mpi} and \citep{matsumoto2007self}, data reconstruction onto the refined grid is done by a linear interpolation, which lowers the numerical diffusion of shock discontinuities, as well as prevents the generation of spurious shocks on cells that abut on coarser ones.
Explicitly, the conservative variables $\mathbf{U}$ in the domain $D^{\ell+1}$ are:

\begin{equation}\label{Linear interpolation}
\mathbf{U}^{\ell+1}_{i,j,k}=\mathbf{U}^{\ell}_{I,J,K}+\overline{\nabla}\left(\mathbf{U}^{\ell}_{I,J,K}\right)\cdot\left( \mathbf{r}^{\ell}_{I,J,K}-\mathbf{r}^{\ell+1}_{i,j,k}\right),
\end{equation}

\noindent where $\mathbf{r}^{\ell+1}_{i,j,k}$ are the cell positions of $D^{\ell+1}$, and $\mathbf{r}^{\ell}_{I,J,K}$ the position of the cell in $D^{\ell}$, where the conservative variables take values $\mathbf{U}^{\ell}_{I,J,K}$.

The approximation of the gradient operator, $\overline{\nabla}$, acting on the arbitrary scalar function $\varphi$, is the one presented in \citep{matsumoto2007self}, given by:

\begin{equation}\label{Nabla minmod}
    \overline{\nabla}\left(\varphi\right)=\begin{pmatrix}
    \mathrm{Minmod}\left(\partial_{x+1/2}\varphi,\partial_{x-1/2}\varphi\right)\\
    \mathrm{Minmod}\left(\partial_{y+1/2}\varphi,\partial_{y-1/2}\varphi\right)\\
    \mathrm{Minmod}\left(\partial_{z+1/2}\varphi,\partial_{z-1/2}\varphi\right)
    \end{pmatrix},
\end{equation}

\noindent and the derivative operation follows the second order accurate centered finite differences: 

\begin{equation}\label{Partial x}
\partial_{x+1/2}\varphi_{i,j,k}=\frac{1}{\Delta x}(\varphi_{i+1,j,k}-\varphi_{i,j,k}),
\end{equation}

\noindent and the Minmod function is given by

\begin{equation}\label{Minmod}
    \mathrm{Minmod}(a,b)=\left\{ \begin{array}{llll}
        a & \mathrm{if }&|a|<|b|,&  ab>0\\
        b & \mathrm{if }&|b|<|a|,&  ab>0\\
        0 &{}&\mathrm{otherwise} 
    \end{array} \right. .
\end{equation}

\noindent Similar formulas arise for partial derivatives along $y$ and $z$.
It is possible to use more robust linear reconstruction formulas, such as MC, or higher order WENO, as long as they preserve overall total variation diminishing properties, which will not affect the overall schemes. However, in the examples of this paper, we have chosen to only use minmod as it is sufficiently precise in regions with big discontinuities, while also being diffusive enough. This helps to prevent the growth of disturbances created by boundaries between refinement levels.

\subsection{Time Evolution and Synchronization.}
\label{subsection:Time Evolution and synchronization}

The AMR strategy we implement generates multiple self similar grids, each one of them evolving independently in time by the scheme \eqref{Flux} and only sharing information through their boundaries at the beginning of the evolution from time $t^n$ to $t^{n+1}$.

The Godunov type of finite volume time evolution used for every grid has the Courant-Friedrichs-Lewy (CFL) stability condition that relates the time step $\Delta t$ with the grid resolution by 

\begin{equation}\label{eq:CFL}
    \Delta t= CFL\times\frac{{\min[\Delta x, \Delta y, \Delta z]}}{\lambda_{max}},
\end{equation}

\noindent where $\lambda_{max}$ is defined as the maximum characteristic speed of the linearized version of equations \eqref{eq: mass conservation}-\eqref{eq: Psi} throughout the whole domain; the definition of this speed in three dimensional problems   reads:

\begin{equation}\label{eq:lambda max}
\lambda_{\max}=\max[|v_x|+C_{fx},|v_y|+C_{fy},|v_z|+C_{fz}]_{i,j,k}
\end{equation}

\noindent where $C_{fs}$ is the maximum magnetosonic speed

\begin{equation}\label{eq:fast magnetosonic speed}
    C_{fs}=\sqrt{\frac{1}{2}\left[ a+b+\sqrt{(a+b)^2-4ab_s}\right]},
\end{equation}

\noindent where $a=\gamma P/\rho$, $b=B^2/\rho$, $b_s=B_s^2/\rho$, and $s=x,y,z$. As stated in \citep{toro2013riemann}, the condition $CFL<1.0$ is required to preserve stability.

Following the indications in \citep{berger1984amr} and \citep{berger1989local}, time step size $\Delta t^{\ell+1}$ of refinement level $\ell+1$ is related with $\Delta t^{\ell}$ of refinement level $\ell$ by 

\begin{equation}\label{time ref}
    \Delta t^{\ell+1}=\frac{1}{r}\Delta t^{\ell};
\end{equation}

\noindent where $r$ is the ratio between cell sizes of  refinement level $l$ and that of level $l+1$. In this code we chose a constant refinement ratio $r=2$.
To obtain \eqref{time ref} using \eqref{eq:CFL} we get $\lambda_{max}$ among all subgrids and select the maximum. Also in this step, following \citep{dedner2002hyperbolic} we define the coefficient $C_h=\lambda_{\max}$ used in equations \eqref{eq: Psi} and \eqref{eq:C_r} of the divergence cleaning method.

\begin{figure}
    \centering
    \includegraphics[scale=0.3]{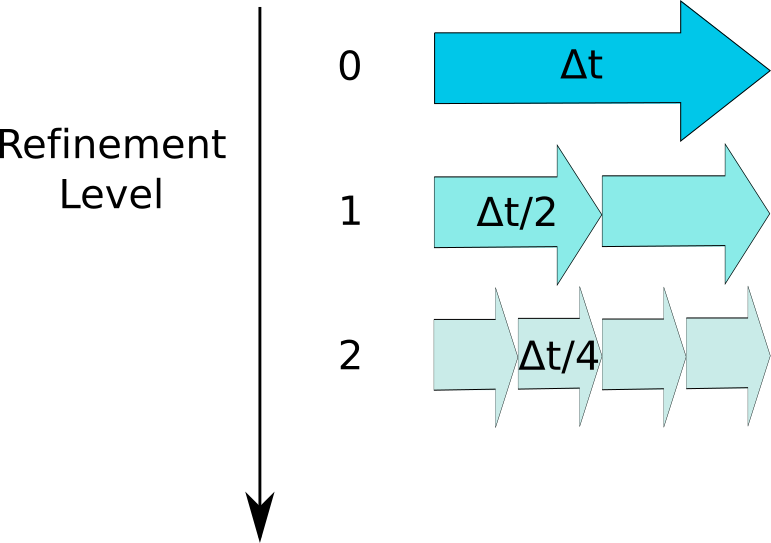}
    \caption{Schematic illustration of time steps applied to each refinement level for grids with $\ell=0,1,2$ to advance from $t^n$ to $t^{n+1}$. A similar scheme holds for a bigger number of refinement levels.}
    \label{fig:Time step}
\end{figure}

A complete evolution cycle of the system from $t^n$ to $t^{n+1}$ is considered as done once all of the grids synchronize at the time $t^{n+1}$. This requires that the domain $D^{\ell}$ must be evolved $2^\ell$ times to catch up with the coarsest domain. Figure \ref{fig:Time step} shows how grids of multiple refinement level advance several times to do so.

In \citep{keppens2003adaptive}, it was found that choosing a non-homogeneous refinement ratio independently of the overall scheme can optimize certain aspects of the simulation, leading to faster computing times. However, our observations indicate that the gain from this approach, especially with a refinement ratio greater than 2, is outweighed by the benefits of using the recursive code structure of the fixed ratio. The recursive approach facilitated our control of proper refinement nesting and hierarchical time stepping.

Various AMR codes prove that in many physical simulations, homogeneous time step is preferred, as seen in the Nirvana code \citep{nirvana}, which employs a single time step, not only for the better handling of simulations with non zero diffusion source terms, but also to realize faster parallelization schemes. In our tests, hierarchical time stepping yielded good results.

\begin{figure}
\centering
\includegraphics[width=6.cm]{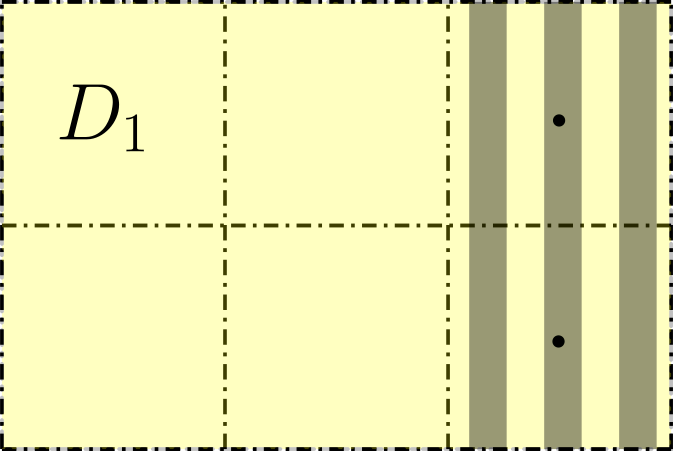}\\\vspace{0.25cm}
\includegraphics[width=6.cm]{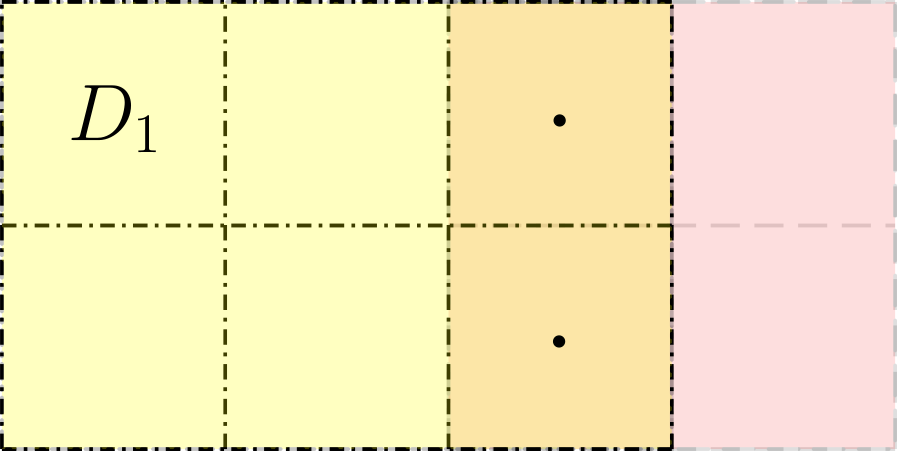}\\\vspace{0.25cm}
\includegraphics[width=6.cm]{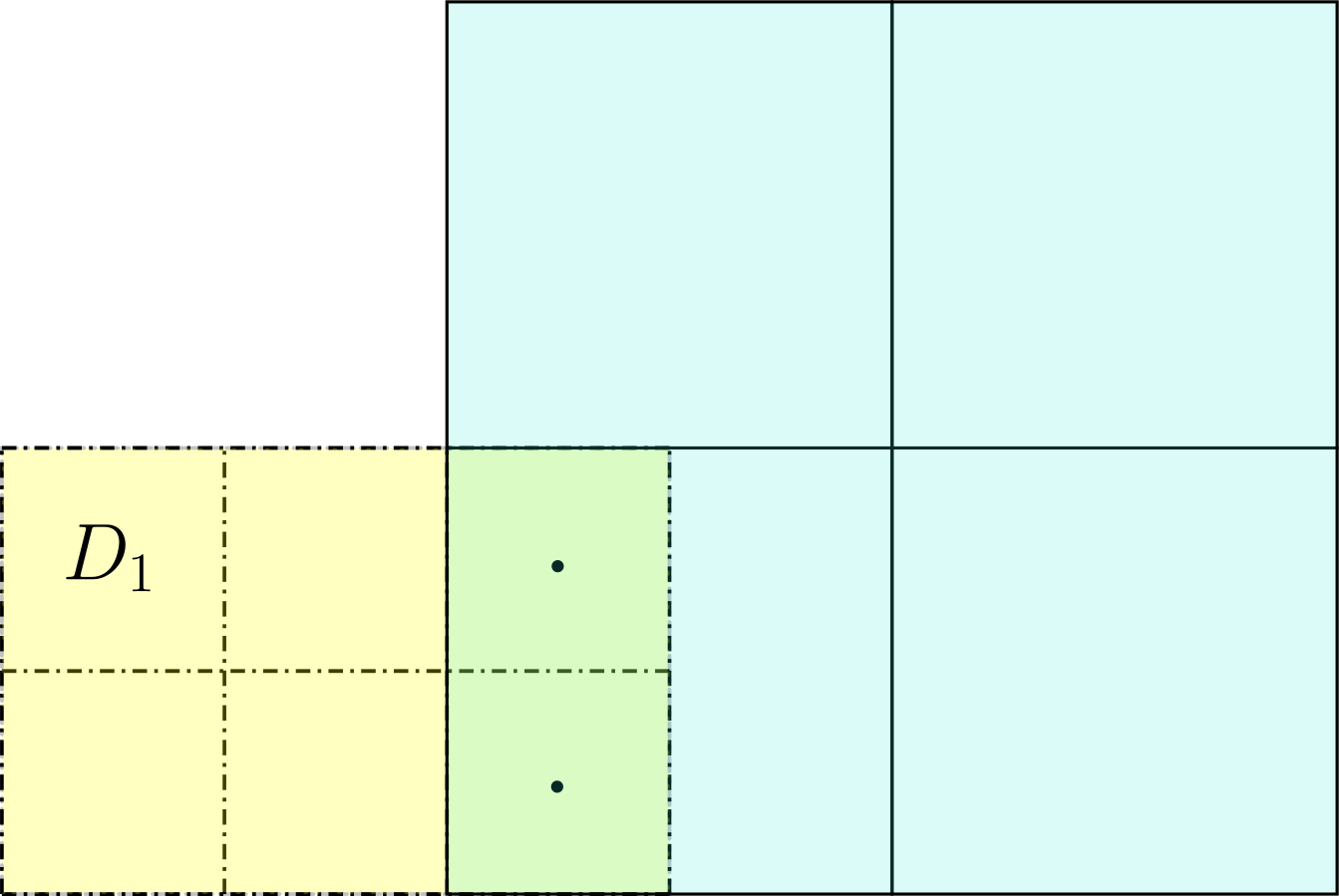}
\caption{\label{fig:Boundaries} 2-dimensional sketch of ghost cell locations of an arbitrary domain $D_1$ according to the type of boundary they share. Ghost cells are marked with a dot in each case. (Top) Physical boundary, (middle) shared boundary with domains of same level of refinement and (bottom) shared boundary with domains of same refinement level.} 
\end{figure}

Except for the boundaries of the physical domain, the evolution of refined domains $D^{\ell}$ starts with the implementation of boundary conditions. This is done by filling \textit{ghost cells} that extend the domain $D^\ell$, filled with appropriate flux values, see e.g. \citep{leveque2002finite}. 
Filling of ghost cells depends on the kind of boundary they are extending. As shown in Figure \ref{fig:Boundaries}, we identify three types of boundaries of domain $D^{\ell}$:

\begin{itemize}
    \item Shared boundaries with another domain of the same refinement level $\ell$.
    \item Shared boundary with domains of lower refinement level, which may be only of level $\ell - 1$ because of how new sub grids are generated.
    \item Shared boundary with the physical boundary.
\end{itemize}

According to these types of boundaries ghost cells are filled in the following way: 

\begin{itemize}
    \item If ghost cells have the same position as cells contained on another same refinement level grid, simply copy data values contained on those cells.
    \item If ghost cells of a domain $D^{\ell}$ are contained on a coarse cell from a domain $D^{\ell-1}$ linearly interpolate the data values.
    \item If ghost cells correspond to a physical boundary, implement the appropriate boundary conditions, such as outflow or rigid, to name a few.
\end{itemize}

In the case of boundaries between consecutive refinement levels, the values of the variables at ghost cells of $D^{\ell}$ are constructed using linear interpolation from the values at cells in  $D^{\ell-1}$ in the same way as equation \eqref{Linear interpolation}, that is 

\begin{equation}\label{eq: CoarseToRef Boundary}
    \mathbf{U}^{\ell}_{i,j,k}=\mathbf{U}^{*\ell }_{I,J,K}+\hat{\nabla}\left(\mathbf{U}^{*\ell-1}_{I,J,K}\right)\cdot\left( \mathbf{r}^{\ell-1}_{I,J,K}-\mathbf{r}^{\ell}_{i,j,k}\right),
\end{equation}
where it is necessary to define the coarse grid values $\mathbf{U}^{*\ell}$, as
\begin{equation}\label{eq:U Coarse timeAverage}
    \mathbf{U}^{*\ell}=\left\{ \begin{array}{ccc}
        \mathbf{U}^{\ell}(t) & \mathrm{if}&\mathrm{synchronized},\\
        \frac{1}{2}\left( \mathbf{U}^{\ell}(t+\Delta t^\ell)+\mathbf{U}^{\ell}(t)\right) &{}&\mathrm{otherwise}
    \end{array} \right.
\end{equation}

\noindent and where the numerical gradient $\hat{\nabla}$ is an upwind  or downwind approximation depending on which face of the coarse level $\ell-1$ grid abuts on a refined one $\ell$. For example, if the boundary perpendicular to the $+\hat{x}$ direction of $D^{\ell-1}$ is shared with the refined domain $D^{\ell}$, then

\begin{equation}\label{eq: Gradient CoarseToRef}
    \hat{\nabla}(\varphi)=\begin{pmatrix}
    \partial_{x-1/2}\varphi\\
    \mathrm{Minmod}\left(\partial_{y+1/2}\varphi,\partial_{y-1/2}\varphi\right)\\
    \mathrm{Minmod}\left(\partial_{z+1/2}\varphi,\partial_{z-1/2}\varphi\right)
    \end{pmatrix},
\end{equation}

\noindent where $\varphi$ is an arbitrary scalar function; if the boundary perpendicular to the $-\hat{x}$ direction is the one being shared, then $\partial_{x+1/2}$ is used instead. A similar formula is used when the faces are perpendicular to $\pm \hat{y}$, $\pm \hat{z}$. 

\subsection{Injection of refined data onto the coarser grid.}
\label{subsec: Injection}

Once the variables in the coarse grid and its refinement are synchronized, the values of the conservative variables in the refined grid must be injected onto the coarse one. This is done by overwriting the data values contained in the coarse cell, with the average of the data values in the refined cells that cover it.

In three dimensions this is done using the formula

\begin{equation}\label{eq: Data Injection}
    \mathbf{U}^{\ell}_{I,J,K}=\frac{1}{2^{3}}\sum_{i=1}^{2}\sum_{j=1}^{2}\sum_{k=1}^{2}\mathbf{U}^{\ell+1}_{i,j,k},
\end{equation}

\noindent where $\mathbf{U}^{\ell+1}_{i,j,k}$ are the data values of the refined cells that cover the coarse cell at position $\mathbf{r}_{I,J,K}$.

\subsection{Flux corrections.}
\label{subsec: Flux Correction}

Finite volume methods are based on the conservation of quantities at a given cell. For example, the conservation of mass equation \eqref{eq: mass conservation} on its finite volume form, equation \eqref{Flux}, at cell $(i,j,k)$  can be written as:

\begin{equation}\label{eq: FV mass conservation}
\begin{split}
  \delta m_{i,j,k}&=\Delta \mathrm V\times \left[\frac{\Delta t}{\Delta x}\left(F^n_{i-1/2,j,k}-F^n_{i+1/2,j,k}\right)\right.\\
    {}&+ \frac{\Delta t}{\Delta y}\left(G^n_{i,j-1/2,k}-G^n_{i,j+1/2,k}\right)\\
    {}&+\left.\frac{\Delta t}{\Delta z} \left(H^n_{i,j,k-1/2} -H^n_{i,j,k+1/2}\right) \right],
\end{split}
\end{equation}

\noindent where $\Delta V=\Delta x \Delta y \Delta z$ is the cell volume, $\delta m_{i,j,k}= \Delta V(\rho_{i,j,k}(t+\Delta t)-\rho_{i,j,k}(t))$ is the total mass gain over the time step $\Delta t$ and $F$=$\rho v^x$, $G$ =$\rho v^y$, $H$=$\rho v^z$ are the numerical fluxes across the faces of the cell whose normal directions are respectively $\pm\hat{x}$, $\pm\hat{y}$, $\pm\hat{z}$.

Equation \eqref{eq: FV mass conservation} establishes that each of the numerical fluxes introduces an amount of mass gain or loss across the faces of a cell. We focus our attention on the face boundary at $x=x_{i+1/2}$ where the mass gain is 
 
\begin{equation}\label{eq: FV mass conservation Right}
     \delta m_{i+1/2,j,k}(t+\Delta t)=-\Delta t\times \Delta y \Delta z F_{i+1/2,j,k}(t).
\end{equation}

\begin{figure}
     \centering
     \includegraphics[scale=0.25]{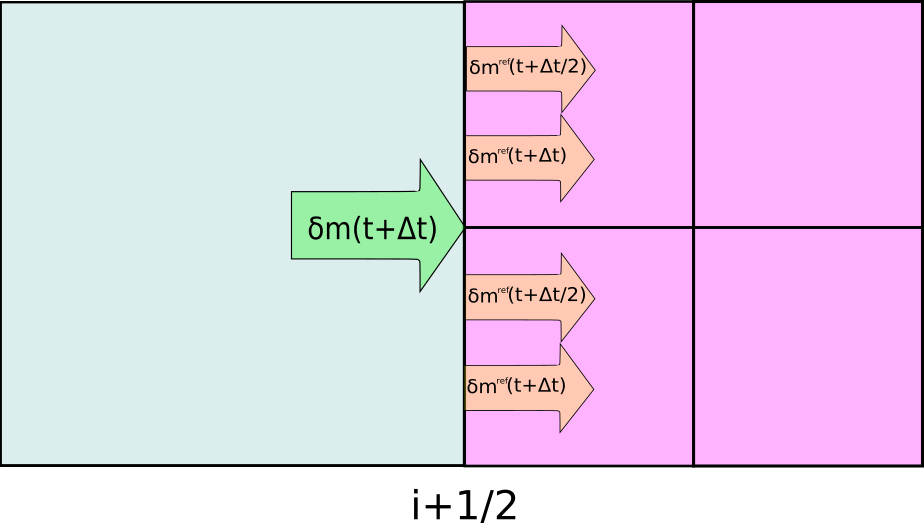}
     \caption{Illustration of a boundary between refinement levels and the mass transfer from a cell of $D^{\ell}$ through its face at $x=x_{i+1/2}$, towards its neighboring cells of $D^{\ell+1}$ during a time step. Notice that $\delta m_{i+1/2,j,k}$ has to be computed at times $t+\Delta t/2$ and at $t+\Delta t$.}
     \label{fig:Flux correction}
\end{figure}

\noindent Now, consider a boundary between two domains of different refinement level. When a coarse grid cell abuts on refined grid cells to the right along $x$, the mass variation of equation \eqref{eq: FV mass conservation Right} must be equal on both sides of the boundary, as in Figure \ref{fig:Flux correction}, then the following relation should hold:
 
 \begin{equation}\label{eq:flux mass eq}
 \begin{split}
     \delta m_{i+1/2,j,k}(t+\Delta t)=\sum_{q}\sum_{r}& \left(\delta m_{p-1/2,q,r}(t+\Delta t/2)\right.\\
     +&\left.\delta m_{p-1/2,q,r}(t+\Delta t)\right),
 \end{split}
 \end{equation}

\noindent where indices $(i,j,k)$ label the cell of $D^{\ell}$ and $(p,q,r)$ the cells of $D^{\ell+1}$ abutting on the coarse cell to the left.

By expressing \eqref{eq:flux mass eq} in terms of  \eqref{eq: FV mass conservation Right}, an equivalence of the numerical fluxes of the cells in refinement level $\ell-1$ and the cells in refinement level $\ell$ across the shared boundary is exposed:
 
\begin{equation}\label{eq: Flux equivalence}
\begin{split}
     -F^{\ell-1}_{i+1/2,j,k}(t)=+\frac{1}{8}\sum_m\sum_n&\left(F^{\ell}_{l-1/2,m.n}(t)\right.\\
     &+\left.F^{\ell}_{l-1/2,m.n}(t+\Delta t/2)\right.),
\end{split}
\end{equation}

\noindent this argument must hold not only for all cell faces abutting on refined grids, but also for all the numerical fluxes of every conservative variables at these faces.

In order to enforce \eqref{eq: Flux equivalence} we define a flux correction given by
 
\begin{equation}\label{eq: Flux correction def}
\begin{array}{ll}
    \delta \mathbf{F}^{\ell-1}_{i+1/2,j,k}(t+\Delta t)=& \mathbf{F}^{\ell-1}_{i+1/2,j,k}(t)-\frac{1}{8}\sum_m\sum_n\left(\mathbf{F}^{\ell}_{l-1/2,m.n}(t)\right.\\
    &+\left.\mathbf{F}^{\ell}_{l-1/2,m.n}(t+\Delta t/2)\right),
\end{array}
\end{equation}
 
\noindent that is subtracted or added, depending on which side of the shared boundary the fine cells are, once the conservative variables on the coarse cell have been updated, that is:
 
\begin{equation}\label{eq: Flux correction}
\begin{array}{ll}
\mathbf{U}^{\ell-1}_{i,j,k}(t+\Delta t)\rightarrow &\mathbf{U}^{\ell-1}_{i,j,k}(t+\Delta t)\\
&-\Delta t\Delta y\Delta z \delta\mathbf{F}^{\ell-1}_{i+1/2,j,k}(t+\Delta t).
\end{array}
\end{equation}

\noindent Flux corrections along the $\hat{y}$ and $\hat{z}$ directions are implemented in a similar way.
 
\section{Code structure}
\label{sec: Code Structure}

In this section we present a general overview of CAFE-AMR flow chart as well as its structure. This code is written in Fortran using f90 standard, it uses the mpich library for parallelization, and the hdf5 library to save data. Other than that, CAFE-AMR contains its own libraries and dependencies.

\subsection{Data structure}
\label{subsec:datastructure}

\begin{figure}
    \centering
    \includegraphics[scale=0.2]{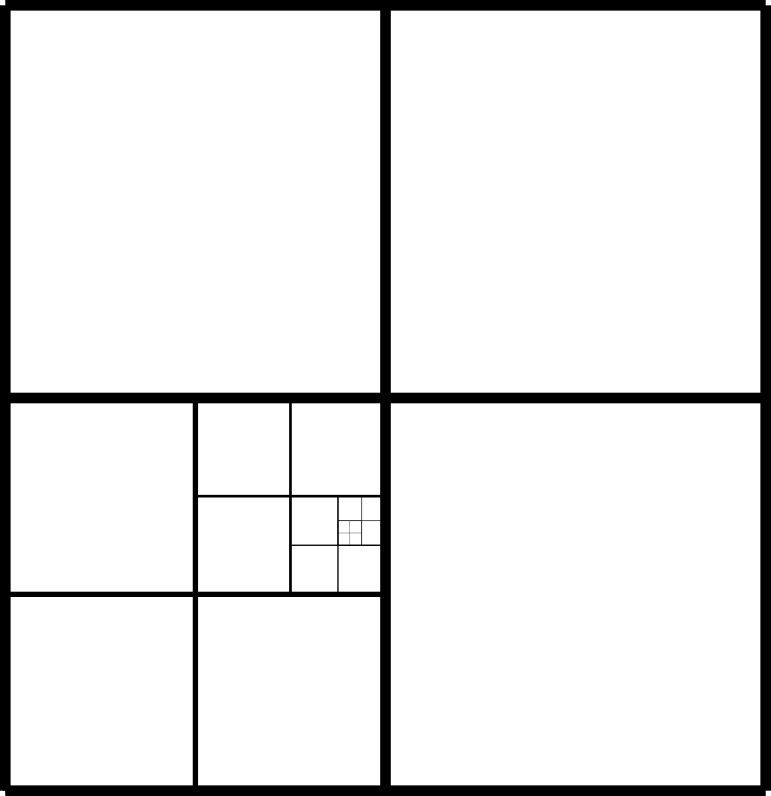}
    \caption{Self similar block structure. In this figure every quadrant of a subdomain can be covered by a block that has the same form and data structure.}
    \label{fig:SelfSimBlock}
\end{figure}

This code works with a self similar block structure, as seen in Figure \ref{fig:SelfSimBlock}, where every block has the following data:

\begin{itemize}
    \item A grid discretized with a regular spacing using $N_x\times N_y\times N_z$ cells plus ghost cells used at the boundaries.
    \item The physical location of the center of the block.
    \item Allocatable arrays. These may contain the values of the conservative variables, or the flux corrections given in \eqref{eq: Flux correction} to the cells abutting on the boundaries of the block.
    \item A refinement level $\ell$ flag that indicates the physical volume of the cells on its grid as well as the time step $\Delta t^{\ell}$.
    \item Pointers to other blocks that share physical boundaries which have the same refinement levels. This blocks are referred to as \textit{neighbors}.
    \item Pointers to other blocks referred to as \textit{children}. These specific blocks cover a quadrant, in two dimensions, or an octant, in three dimensions, contained inside the original block. 
    \item Pointer to another block which has a portion of its domain covered by this block. This block is referred to as \textit{father}.
    \item Several logical flags. They have various uses, such as defining boundary conditions and deciding whether memory will be used, to name a few.
\end{itemize}

\begin{figure}
    \centering
    \includegraphics[scale=0.15]{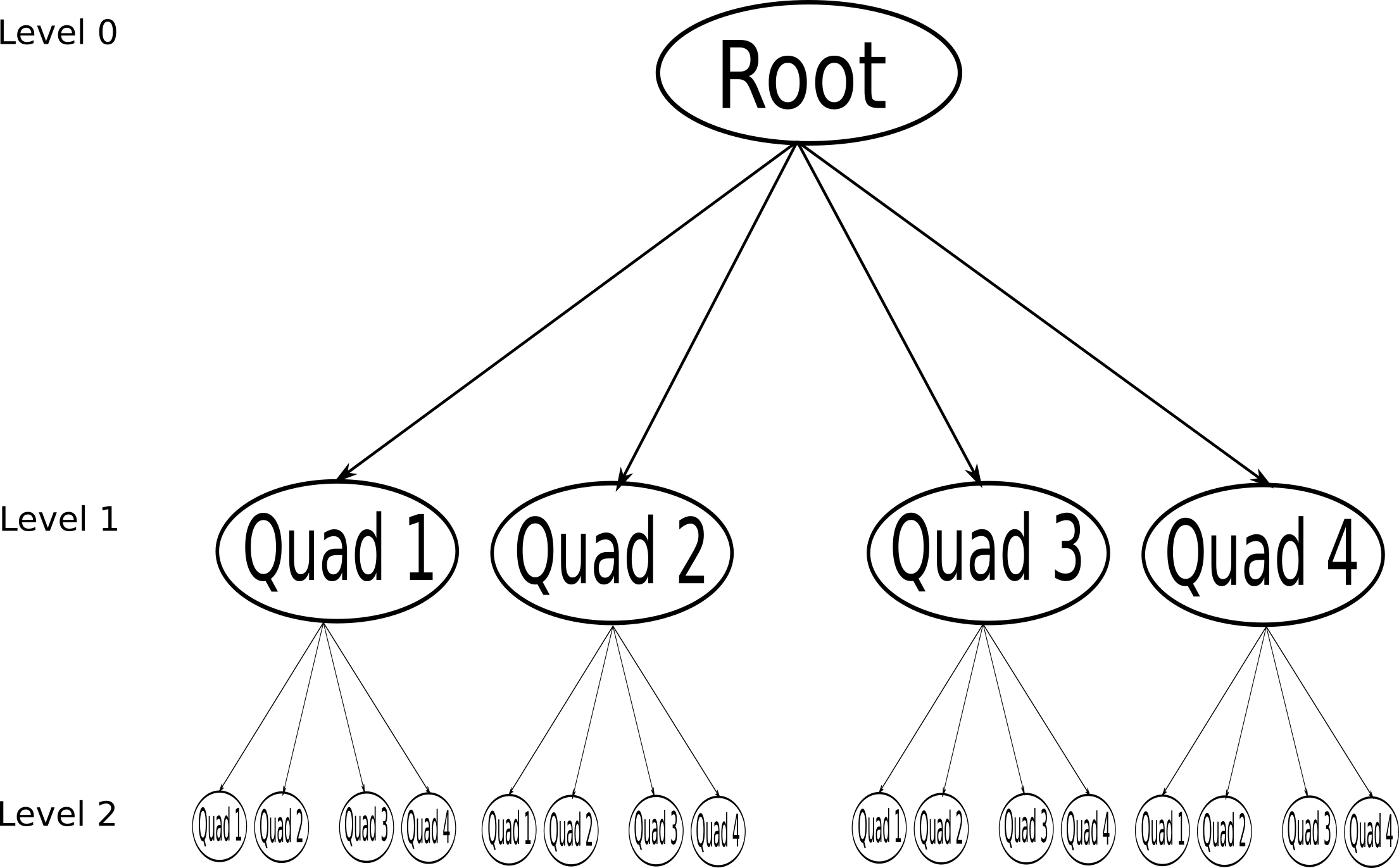}
    \caption{QuadTree data structure used to order data blocks of a two dimensional problem with two refinement levels. Every leaf of the three represents a block of data.}
    \label{fig: QuadTree}
\end{figure}

All the blocks are ordered hierarchically in a QuadTree, in two dimensional problems, and an OcTree in three dimensional problems. In Figure \ref{fig: QuadTree} a QuadTree is represented. In this data hierarchy, the refinement of the domain is generated in the following way. First the blocks are analyzed by quadrants:

\begin{enumerate}
    \item Every time a cell of a given quadrant satisfies the refinement criteria \eqref{Rcriteria}, the whole quadrant is flagged to be refined.
    \item If a Block's quadrant associated child has children of its own, the quadrant is flagged to be refined.
    \item If a quadrant is flagged to be refined, all of its neighbors are also refined, even if they correspond to a different block.
\end{enumerate}

\noindent Then all of the flagged quadrants are covered by newly generated data blocks which join the quadtree as children of the block they cover. Later on, all of the unflagged quadrants have their corresponding children deallocated.

This whole procedure is done from the highest refinement level to the base level so that the blocks are properly nested, according the scheme in subsection \ref{subsec: Selecting a region to refine}. Once the refinement process has ended and all data have been interpolated, the whole neighbor pointer structure is updated. Octrees are refined in the same way.

Finally, if a data block is flagged to abut on a coarser one, a flux correction array is allocated. This array adds on the corrections in order to calculate the terms of Eq. \eqref{eq: Flux correction}. Flags that indicate whether coarse data will be interpolated to fill ghost cells in children blocks are also activated  at the end.

\subsection{Time evolution}

As the data structure suggests, time evolution is done recursively beginning on all the blocks of a given base level to an user specified last level of refinement. The time evolution control flow that holds synchronization of all blocks is the following:

\begin{enumerate}
    \item Apply physical boundary conditions at refinement level $\ell$.
    \item Copy data values from same level neighbors that share a boundary into ghost cells. 
    \item Fill ghost cells values to level $\ell+1$ blocks abutting level $\ell$ blocks by interpolation in a synchronous way following equation \eqref{eq:U Coarse timeAverage}.
    \item Advance level $\ell$ blocks a time-step $\Delta t^\ell$.
    \item Advance \textit{recursively} $\ell+1$ blocks a time-step $\Delta t^{\ell}/2$.
    \item  Fill ghost cell values to level $\ell+1$ blocks abutting level $\ell$ blocks by interpolation in an asynchronous way following equation \eqref{eq:U Coarse timeAverage}.
    \item Advance \textit{recursively} $\ell+1$ blocks a time-step $\Delta t^{\ell}/2$.
    \item Inject values of children blocks into the corresponding region they cover on level $\ell$ blocks using equation \eqref{eq: Data Injection}.
    \item Apply numerical flux corrections on level $\ell$ blocks following \eqref{eq: Flux correction}. 
\end{enumerate}

Flux corrections are implemented by first collecting in allocatable arrays,  both, the cells abutting on $\ell - 1 $ grids and the corresponding cells on coarse blocks, afterwards these data are used to construct the terms $\delta F^{\ell-1}$ of equation \eqref{eq: Flux correction def}.





\section{Tests of the code}
\label{sec:tests}

In this section we present a set of tests that illustrate the ability of the code to handle different problems and the well functioning of the AMR.


\subsection{2D Kelvin-Helmholtz instability}

Kelvin-Helmholtz instabilities are standard tests for hydrodynamical codes, and in the context of Solar Physics these can occur at the boundaries between different regions of the solar wind with varying velocities or densities as well as in small scale solar jets \citep{FedunReview}. As shown in \citep{mishin2016kelvin}. These instabilities can lead to the formation of turbulent structures and the mixing of plasma, which can affect the propagation and properties of the solar wind. 
This is a MHD test, in which the discontinuities in counter-flow regions develop spiral-like shapes on, that are a challenge for an AMR method to follow. 

\begin{figure}
\begin{center}
\includegraphics[width=6.1cm]{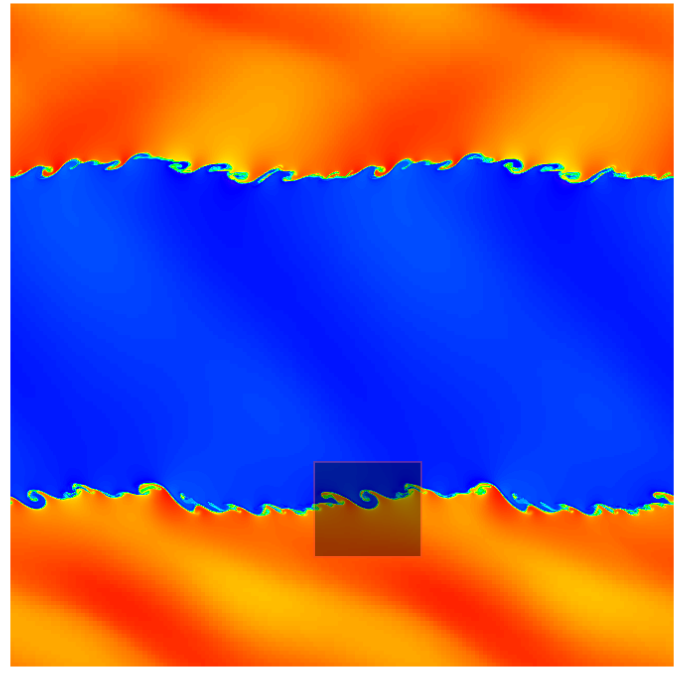}
\includegraphics[width=6.cm]{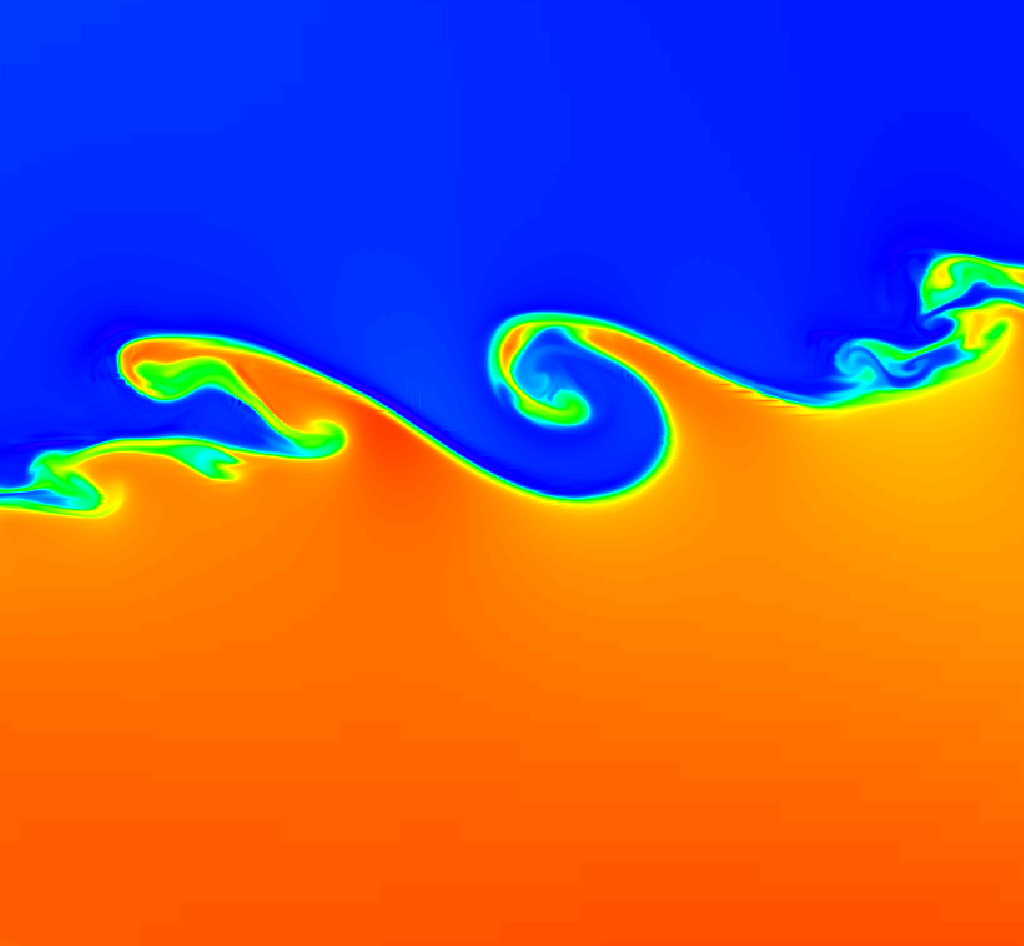}
\caption{\label{fig:BKH5RL_t025} Snapshot of the density for the 2D Kelvin-Helmholtz instability at $t=0.24$. At the top we show the whole domain, with a gray area that we zoom in and show at the bottom.}
\end{center}
\end{figure}

The initial conditions for this test are set to:

{\small 
\begin{eqnarray}
(\rho,p,v^x,v^y,B^x,B^y) &=& \left\{
\begin{array}{ll}
(1,2.5,-0.5,0,0.2,0),     &~~ |y| < 0.52\\
(2,2.5,0.5,0,0.2,0),     &~~ |y| \ge 0.25
\end{array}
\right.\label{eq:IDKelvinHelmholtzB}
\end{eqnarray}
}

\noindent with adiabatic index $\gamma=1.4$. In order to trigger the instability, the velocity components are perturbed as follows $v^x=v^x+\delta$ and $v^y=v^y+\delta$, where $\delta=0.1\cos(4\pi x)\sin(4\pi y)$.

This test is defined in the periodic domain $[-0.5,0.5]\times[-0.5,0.5]$, with a base resolution of $160\times160$ cells, time step resolution given by a CFL factor $0.75$. For this test we use the Roe flux formula and the CTU time stepping scheme. We present three simulations with different refinement criteria as well as different number of refinement levels.

\begin{figure}
\begin{center}
\includegraphics[width=6.cm]{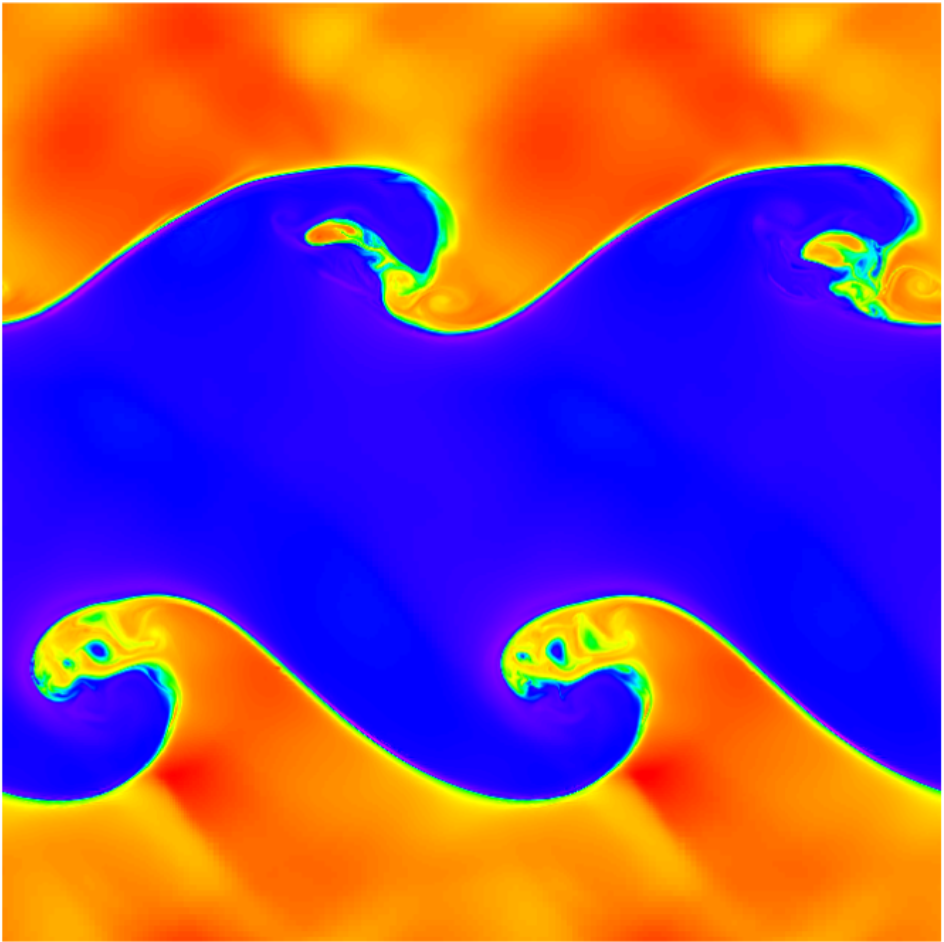}\vspace{0.2cm}
\includegraphics[width=6.cm]{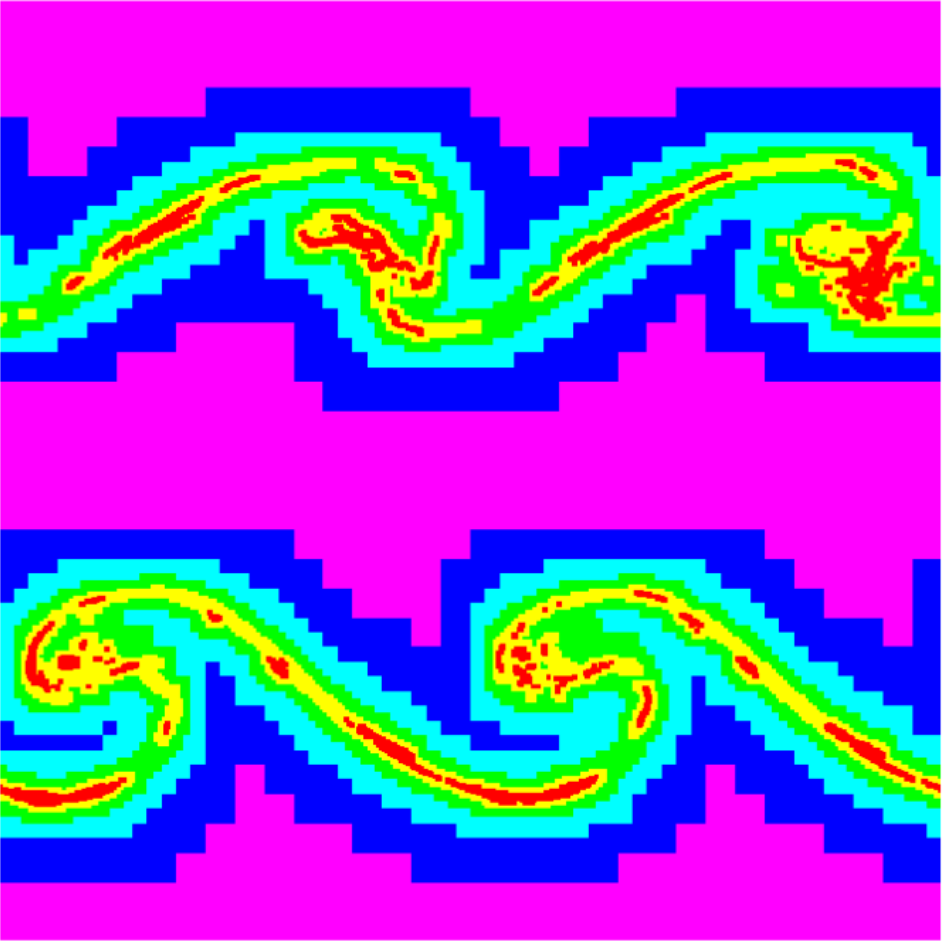}
\end{center}
\caption{\label{fig:BKH5RL_t01} Snapshot of the 2D Kelvin-Helmholtz instability at $t=1.0$. At the top we show the density and at the bottom the grid with its various refinement levels.}
\end{figure}

In a first simulation we use criterion (\ref{ref criteria}) with a threshold value of $\chi_{r}=0.1$ using the density $\rho$ as the evaluating function in (\ref{Rcriteria}). We use five refinement levels, which gives an equivalent resolution of $5120\times5120$ cells. In Figure \ref{fig:BKH5RL_t025} we show a snapshot along with a zoomed region at the interface region, the addition of resolution across the interface manages to develop spiral like shapes in small spatial scales. In Figure \ref{fig:BKH5RL_t01} we illustrate how the AMR method follows the discontinuity even with a slim refined domain while also maintaining the overall structure of the less dynamical regions.

\begin{figure}
\includegraphics[width=4.cm]{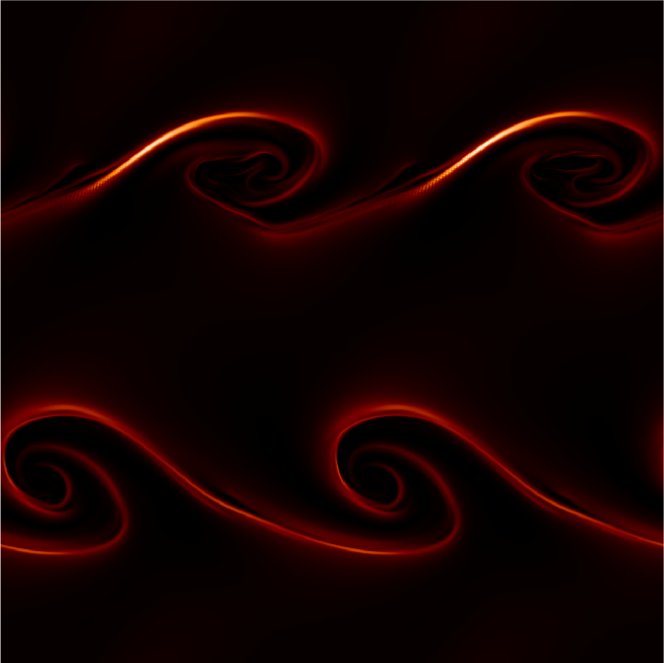}
\includegraphics[width=4.cm]{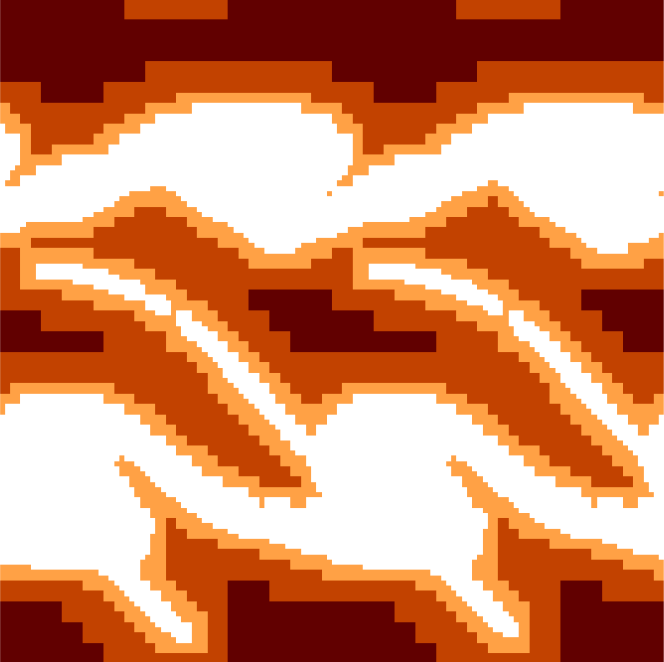}\\
\includegraphics[width=4.cm]{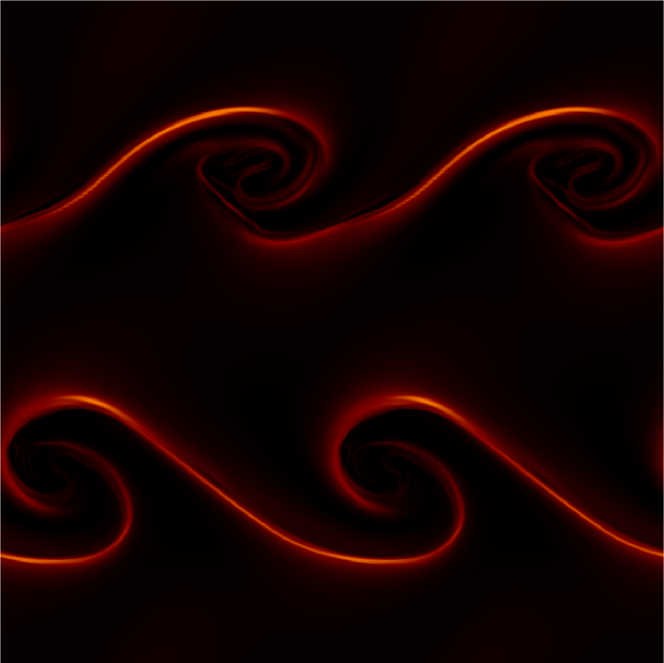}
\includegraphics[width=4.cm]{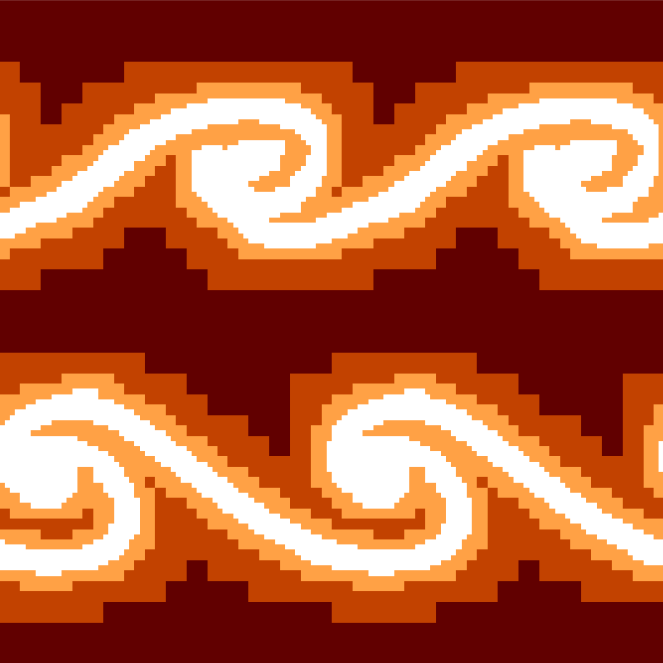}
\caption{\label{fig:BKH4RLChi20} 
Snapshot of the magnetic pressure (left) and the refinement structure (right) of the 2D Kelvin-Helmholtz instability with three refinement levels at $t=1.0$. At the top we show the case with criterion (\ref{RcritMio}) and threshold parameter $\chi_{r}=10$, whereas at the bottom we show the result obtained using criterion (\ref{ref criteria}) with threshold parameter $\chi_r=0.05$.}
\end{figure}

In Figure \ref{fig:BKH4RLChi20} we show the results of a second simulation where we use three refinement levels, giving an equivalent resolution of $1024\times 1024$ cells. In this case, we compare the use of refinement criterion (\ref{RcritMio}) with $\chi_r=10$ and criterion (\ref{ref criteria}) with $\chi_r=0.05$, acting on the density $\rho$. In the refinement structure, it is shown that, since the criterion (\ref{RcritMio}) uses all of the conservative variables as evaluating functions to set the refinement regions, the refined grids are set to cover all of the mixing regions whilst in criterion (\ref{ref criteria}) the refinements are located in smaller regions. This results in a difference on small scale structure of the simulation. It is important to keep in mind this in order to define appropriate refinement strategies.

\subsection{2D Rayleigh-Taylor Instability}

In presence of acceleration, 
when a fluid in low and high density phases, with the low density supporting the weight of the high density phase, initially in hydrostatic equilibrium, small perturbations give rise to the hydrodynamic Rayleigh-Taylor instability.

The analysis of this type of instability is used to learn how an initial linear perturbation develops a complex non-linear behavior. Rayleigh-Taylor instabilities in solar physics serve to gain insights into the physical processes responsible for solar eruptions, energy release, and the dynamics of the Sun's atmosphere, (see e.g. \cite{jenkins2022resolving}). At first, in the linear regime, the growth rate of these instabilities depends on the wave number and frequency of the perturbation, for example the amplitude grows faster for shorter wavelengths. After the perturbation evolves into the nonlinear regime, bubbles from the low density region rise and drops from the high density shell fall, crossing the initial interface surface and the dynamics leads to the mixture of the two initial phases. Its complexity is a reason why this is a challenging test of fluid dynamics codes.

In the case of the MHD, the fluid is considered a magnetized ideal gas that responds to a magnetic field, that changes the evolution of the instability (see e.g. \citep{stone2007nonlinear}). In this test the magnetic field is set parallel to the interface between the two fluid states and inhibits the growth rate of the instability. In order to show that CAFE-AMR can handle this problem, we present the two dimensional Rayleigh-Taylor problem.

We solve the equations on the domain $x\in [-0.5,0.5]$, $y\in[-1.5,0.5]$, and define the interface as the line $y=0$. Initial conditions assume the density in lower and higher parts of the domain are $\rho_L=1$ and $\rho_H=4\rho_L$ respectively, the gravity constant is set to $g=-1.0$ and the system is in hydrostatic equilibrium, consistently with the pressure

\begin{equation}\label{eq:RT_insta Pressure}
    p(y)=\frac{100}{\gamma}-\rho g y,
\end{equation}

\noindent which in turn defines the speed of sound to $c_s=10$ and a crossing time of $t=0.1$ across the interface.
\begin{figure}
    \centering
    \includegraphics[width=2.5cm]{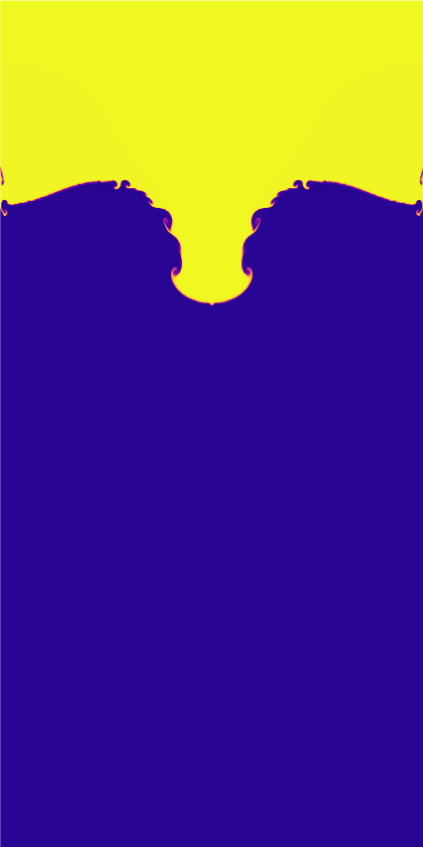}
    \includegraphics[width=2.5cm]{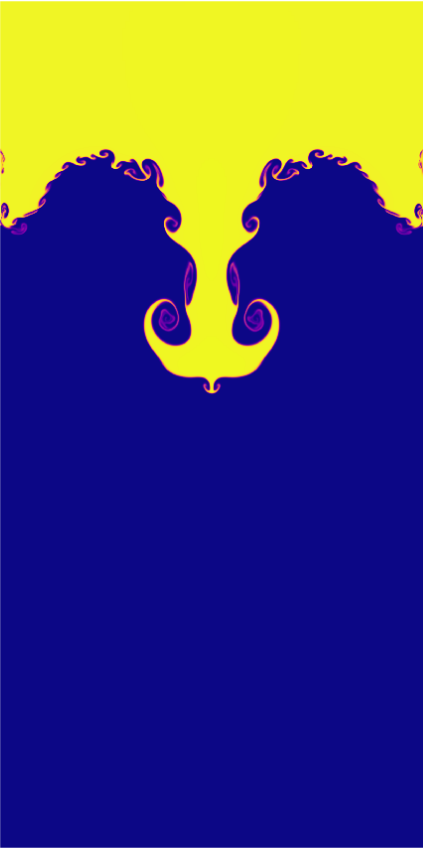}
    \includegraphics[width=2.5cm]{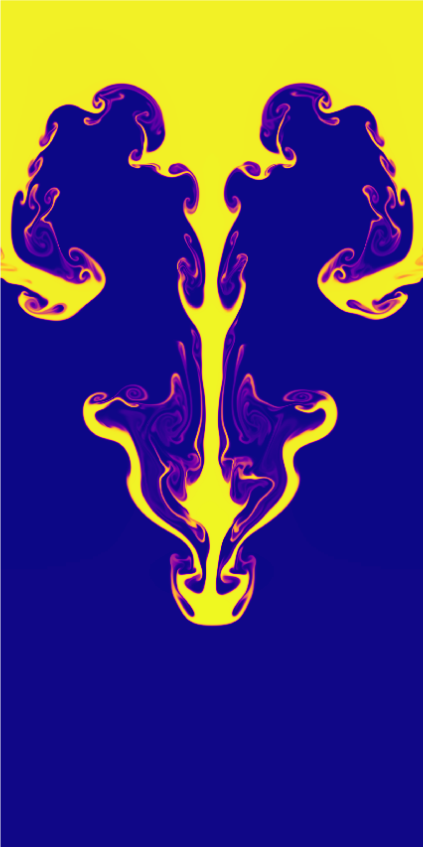}\\
    \includegraphics[width=2.5cm]{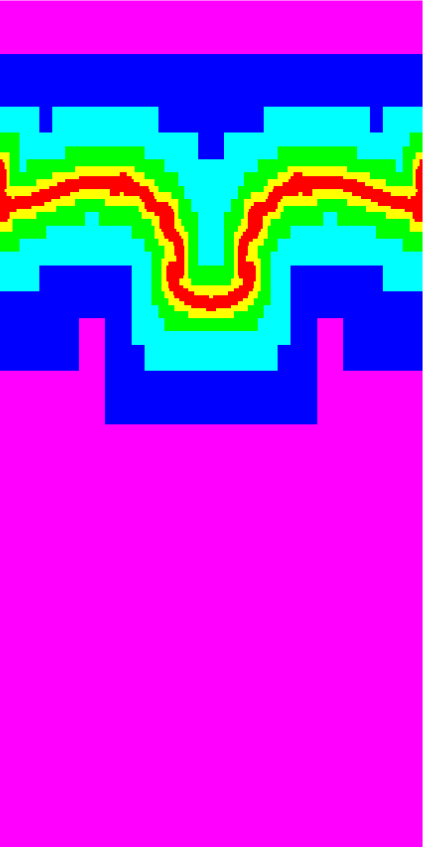}
    \includegraphics[width=2.5cm]{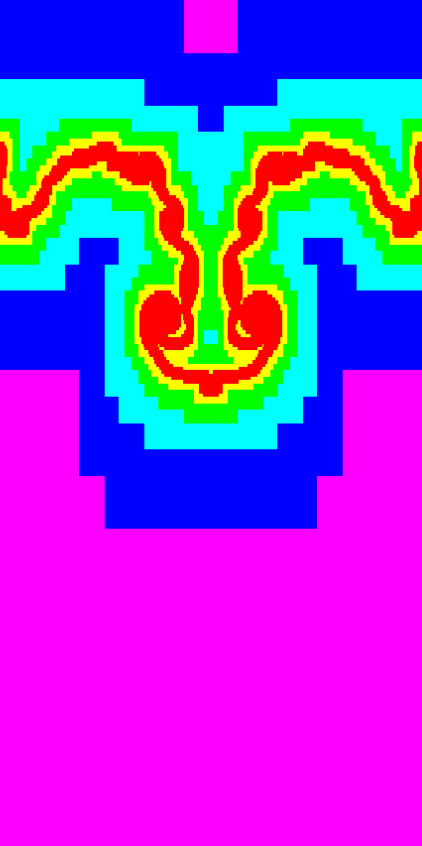}
    \includegraphics[width=2.5cm]{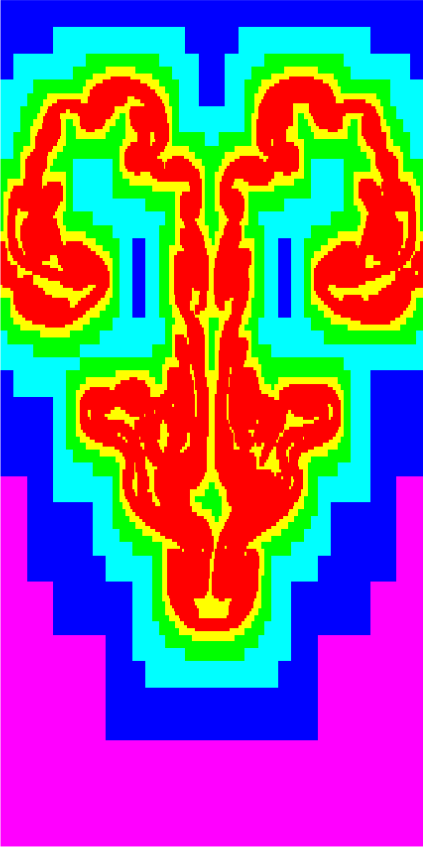}     
    \caption{Snapshots of the density (top) and refinement structure (bottom) of the RT instability in the hydrodynamic limit at times $t=1.5$, $2.0$ and $3.0$, from left to right.}
    \label{fig:2DRT insta hydro} 
\end{figure}

We use periodic boundary conditions along the $x-$direction and fixed conditions at the top and bottom of the domain. The base resolution uses $64\times 128$ cells and we use five refinement levels, which results in an equivalent resolution of $2048\times 4096$ cells with the finest resolution. The refinement criterion used is that in Eq. \eqref{ref criteria} using hydrodynamic pressure, density and the $\hat{y}$ component of the velocity as evaluating functions together with a threshold parameter $\chi_r=0.1$. Numerical fluxes are constructed using the Roe flux formula and the minmod limiter. Time resolution is given by a CFL factor of $0.75$.

\begin{figure}
    \centering
    \includegraphics[width=2.5cm]{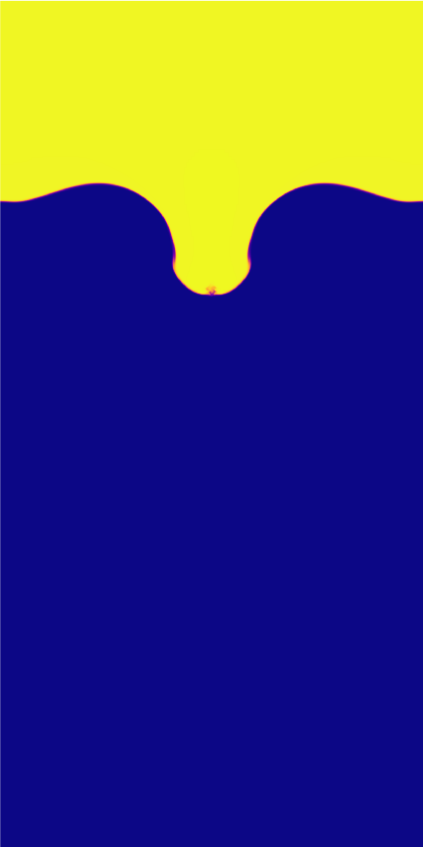}
    \includegraphics[width=2.5cm]{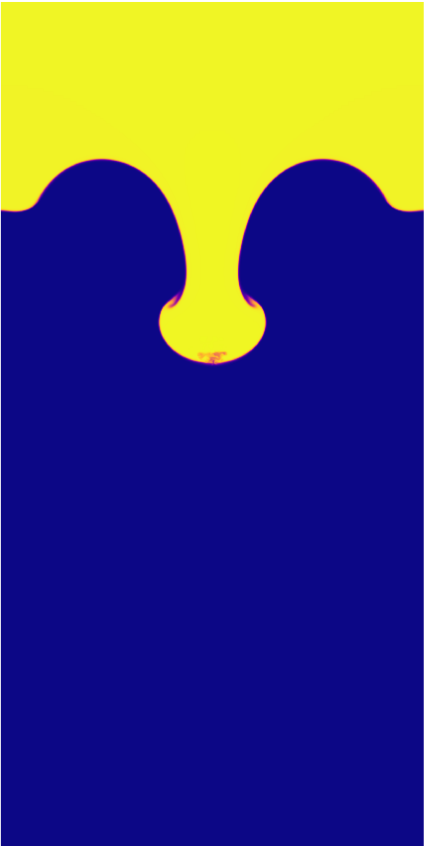}
    \includegraphics[width=2.5cm]{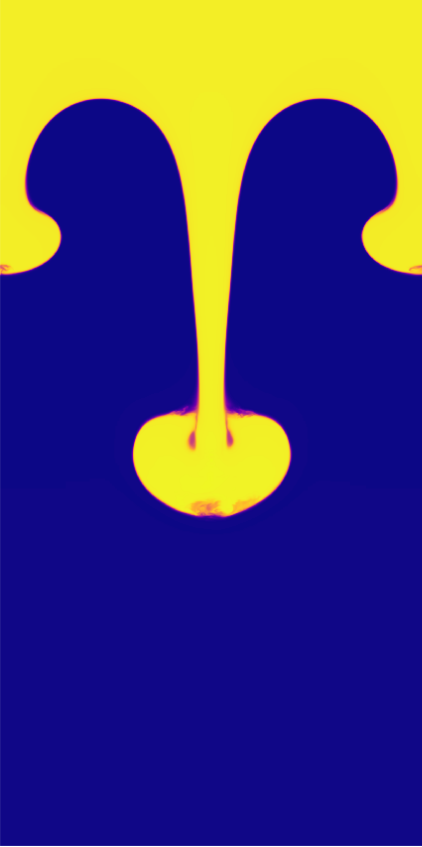}\\
    \includegraphics[width=2.5cm]{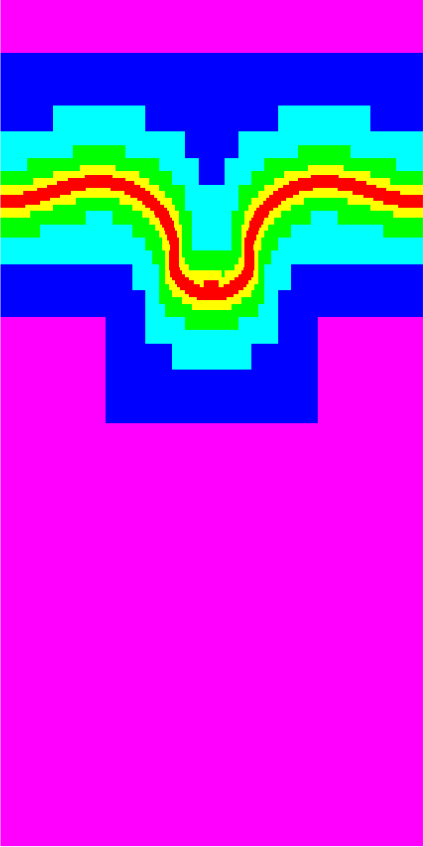}
    \includegraphics[width=2.5cm]{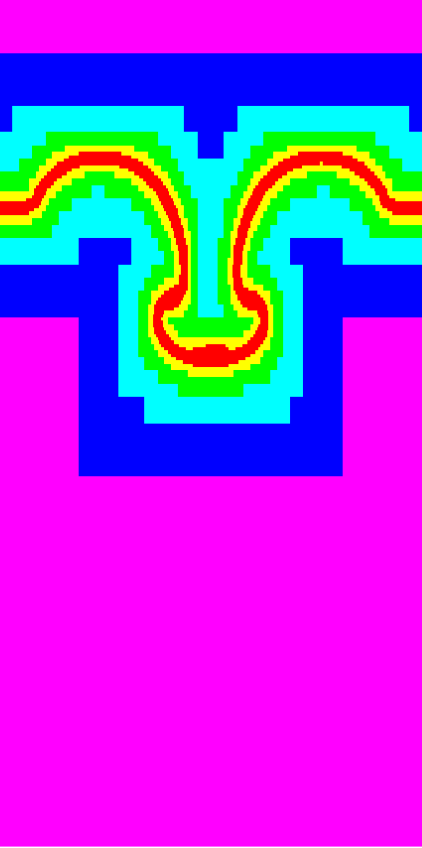}
    \includegraphics[width=2.5cm]{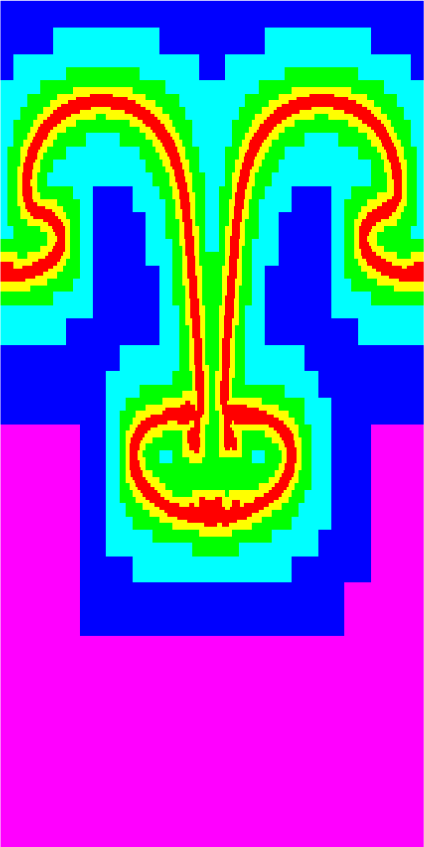}      
     \caption{Snapshots of the density (top) and refinement structure (bottom) of the RT instability in the slightly magnetized case at times $t=1.5$, $2.0$ and $3.0$, from left to right.}
    \label{fig:2DRT insta B}
\end{figure}

\begin{figure}
    \centering
    \includegraphics[scale=0.4]{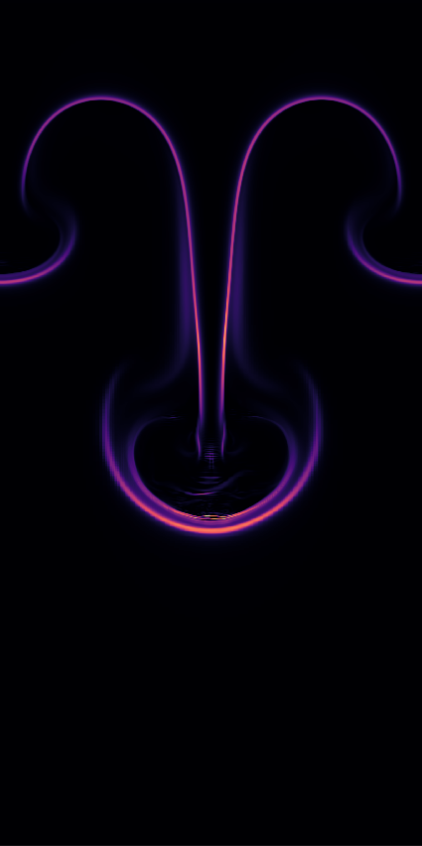}
    \caption{Snapshot of the magnetic pressure at $t=3.0$ of the RT instability test in the slightly magnetized case.}
    \label{fig:2DRT insta Bpr}
\end{figure}

The perturbation that triggers the instability is added to the velocity field at the interface line, assuming the profile

\begin{equation}\label{eq: RT velocity Perturbation}
    v_y(x,y)=-\frac{\mathrm{Exp}[-(5.0 x)^2]}{10\mathrm{cosh}[(10 y)^2]},
\end{equation}

\noindent which is the 2D  version of the instability presented in \citep{mignone2011pluto};  we also define the magnetic field to be constant along the $\hat{x}$ direction,  $\mathbf{B}=(B_x,0,0)$, where  

\begin{equation}\label{eq: RT magnetic field x}
    B_x=\alpha\sqrt{(\rho_H-\rho_L)|g|}. 
\end{equation}

\noindent In our test we present two cases: unmagnetized $\alpha=0$, slightly magnetized $\alpha=0.06$.

\begin{figure}
    \centering
    \includegraphics[width=7.8cm]{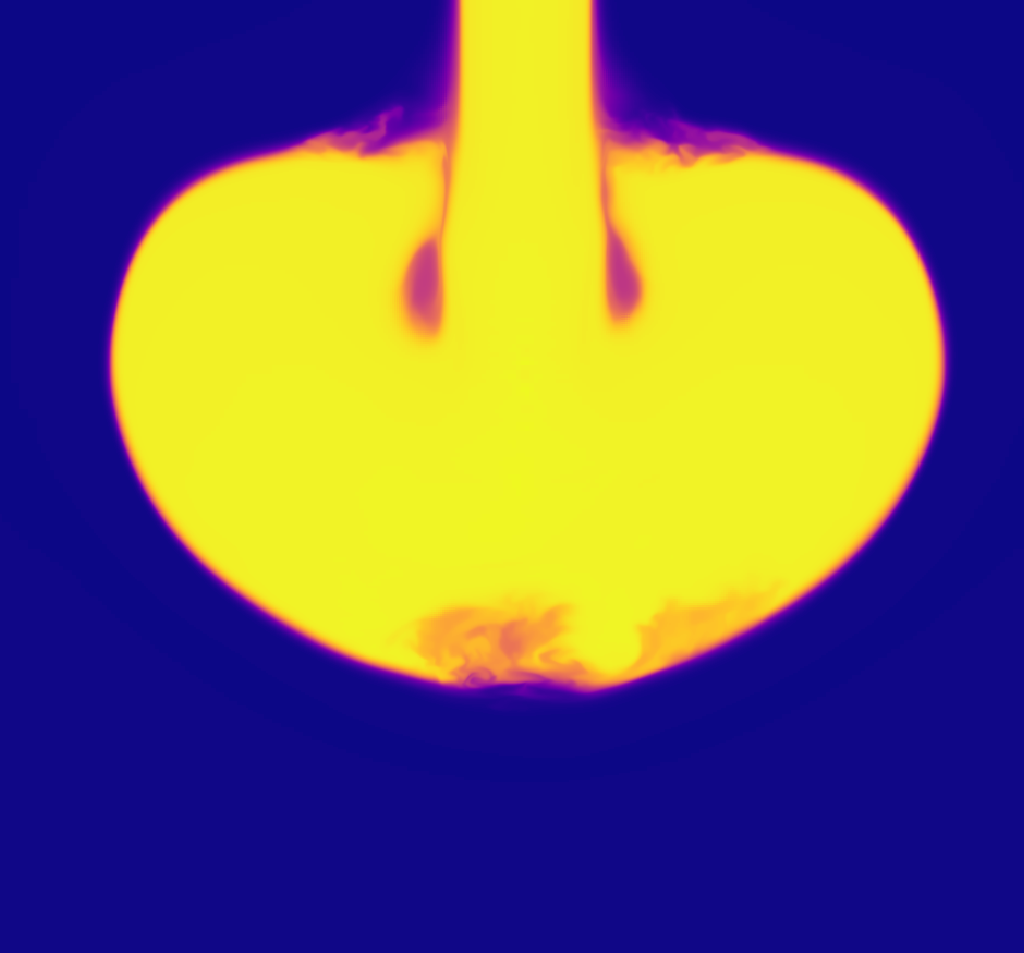}\\
    \includegraphics[width=4.cm]{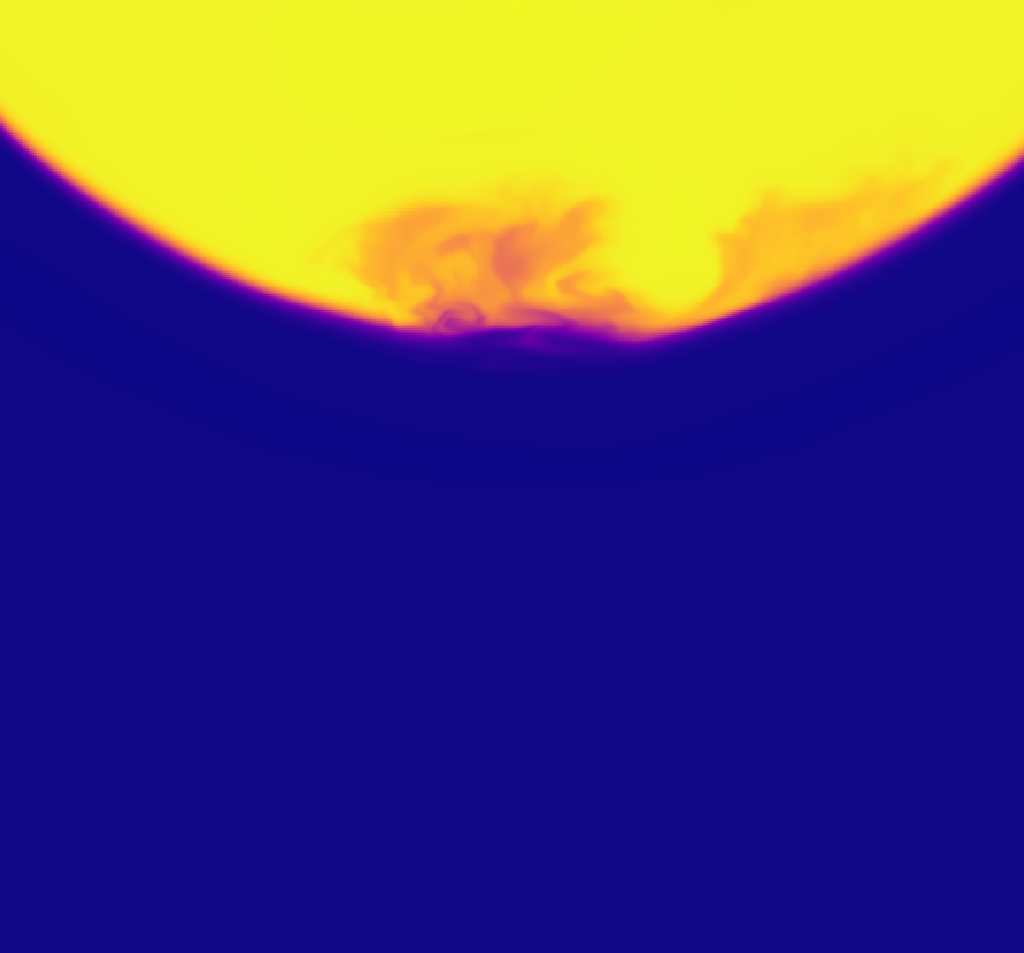}
    \includegraphics[width=4.cm]{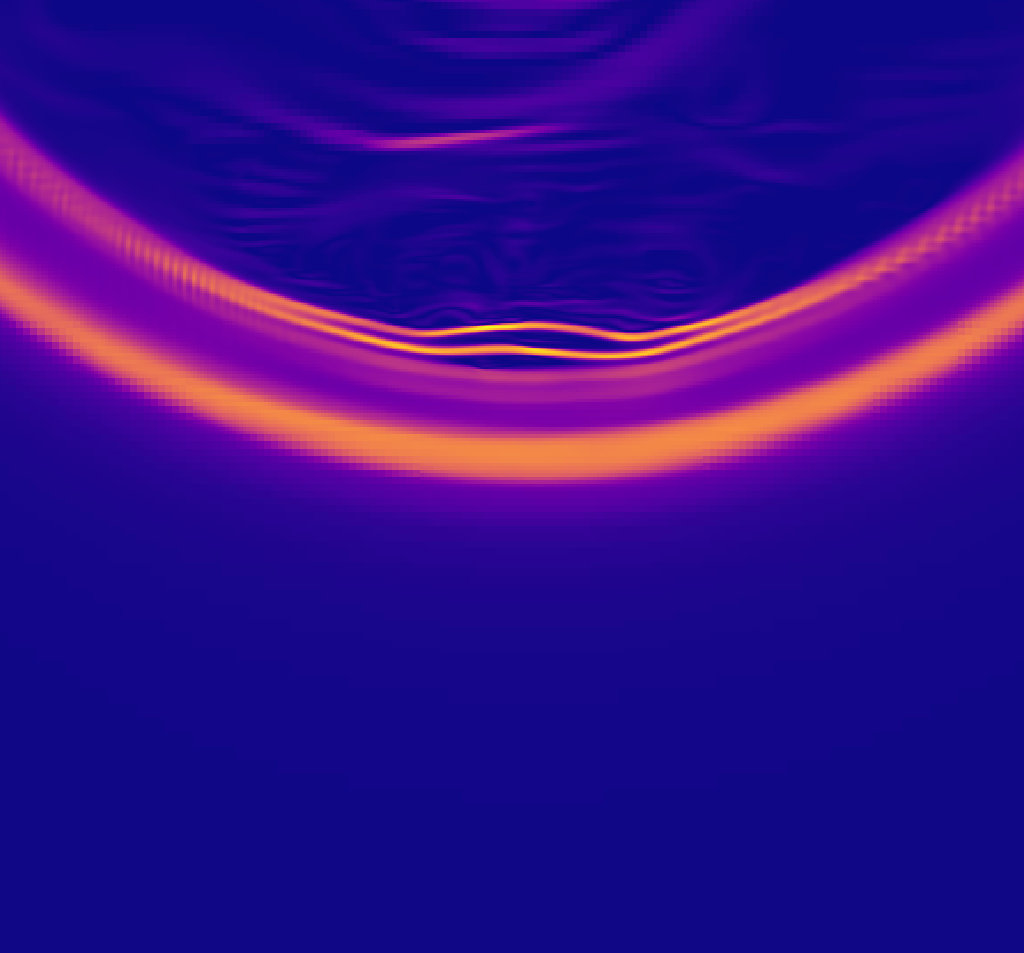}
    \caption{Zoom in of the Rayleigh-Taylor instability test at time $t=3.0$, in the slightly magnetized case. At the top we show the density profile of the lowest region where the interface is located. At the bottom there is a comparison of the density (left) and the magnetic pressure (right) in the turbulent region of the interface.}
    \label{fig:2DRT Insta BZoom}
\end{figure}

Figure \ref{fig:2DRT insta hydro} presents the evolution of the density and refinement structure profiles of the instability at $t=1.5,2.0,3.0$ for the unmagnetized case. As the denser shell begins to fall along the vertical direction, and due to the difference in velocities of the low and high density regions, the flow develops Kelvin-Helmholtz instabilities which in turn develop further into a highly structured interface at latter times. The refinement strategy selected on this case adequately tracks  the overall structure of the interface.

Figure \ref{fig:2DRT insta B} presents the density and refinement structure profiles in the slightly magnetized case $\alpha=0.06$, the Kelvin-Helmholtz instabilities are suppressed because the topology of the field lines interrupts the torsion of the fluid interface, the magnetic field works against gravity and helps the denser shell to remain in the upper part which also stops the generation of KH-instabilities. In Figure \ref{fig:2DRT insta Bpr} we show that the magnetic pressure is acting overall only at the interface surface. In Figure \ref{fig:2DRT Insta BZoom} we show that because of the augmented precision, we can capture at the lowest region of the interface there is a small scale variation in both density and magnetic pressure.

\subsection{Localized Resistivity Current sheet}

Solar flares are modelled as events generated magnetic reconnection (see e.g. \citep{priest2014magnetohydrodynamics}). In \citep{yokoyama2001magnetohydrodynamic} two dimensional MHD simulations were employed to show the effect that localized magnetic resistivity has at triggering this process. The AMR strategy used to approach such simulations is advantageous as locality of phenomena is important. 
With this in mind, we present the localized resistivity Current Sheet (LRCS) test. This test  models the generation of flares in a  stratified density profile were a current sheet magnetic field is developed with the presence of a localized resistivity with Gaussian profile. The simulation is carried out on a numerical domain of $(x,y)\in [-10,10][0,20]$; fixed boundary conditions at the bottom boundary and outflow boundary conditions on the other sides of the domain; we use an initial resolution of $80\times80$ cells with four refinement levels giving an equivalent resolution of $1280\times 1280$; the refinement criterion used is \eqref{RcritMio} with a threshold of $\chi_r=1$; numerical fluxes  are constructed using the HLLE formulas with the second order CTU time stepping. A $CFL=0.01$ factor was chosen so that time steps capture difussion time scales adequately. 

Initial data is defined as follows. Density has the stratified profile

\begin{equation}\label{eq:Flare density}
    \rho(y)=\rho_{chr}+\frac{1}{2}(\rho_{cor}-\rho_{chr})\tanh{[(y-h_{tr})/w_{tr}+1]},
\end{equation}

\noindent where $\rho_{cor}=1$, $\rho_{chr}=\rho_{cor}\times10^5$, $h_{tr}=1$, $w_{tr}=0.2$.  The magnetic field is defined as

\begin{eqnarray}\label{eq:Flare B}
    B_x&=&0,\\
    B_y&=&B_0 \tanh(x/\omega),\\
    B_z&=&B_0 \cosh(x/\omega),
\end{eqnarray}

\noindent where $B_0=1$ and $\omega=0.5$; the localized resistivity function is the same as in \citep{takasao2015magnetohydrodynamic}, that is 

\begin{equation}\label{eq:Flare Eta}
    \eta(x,y)=\eta_0 \exp[-(x^2+(y-h_\eta)^2)/\omega_\eta^2],
\end{equation}

\noindent where $\omega_\eta=0.2$, $h_\eta=6$ and $\eta_0=1$.
Pressure is initially set to $P=1/\gamma$ and the whole system is initially set at rest. 

\begin{figure}
\centering
\includegraphics[width=7.cm]{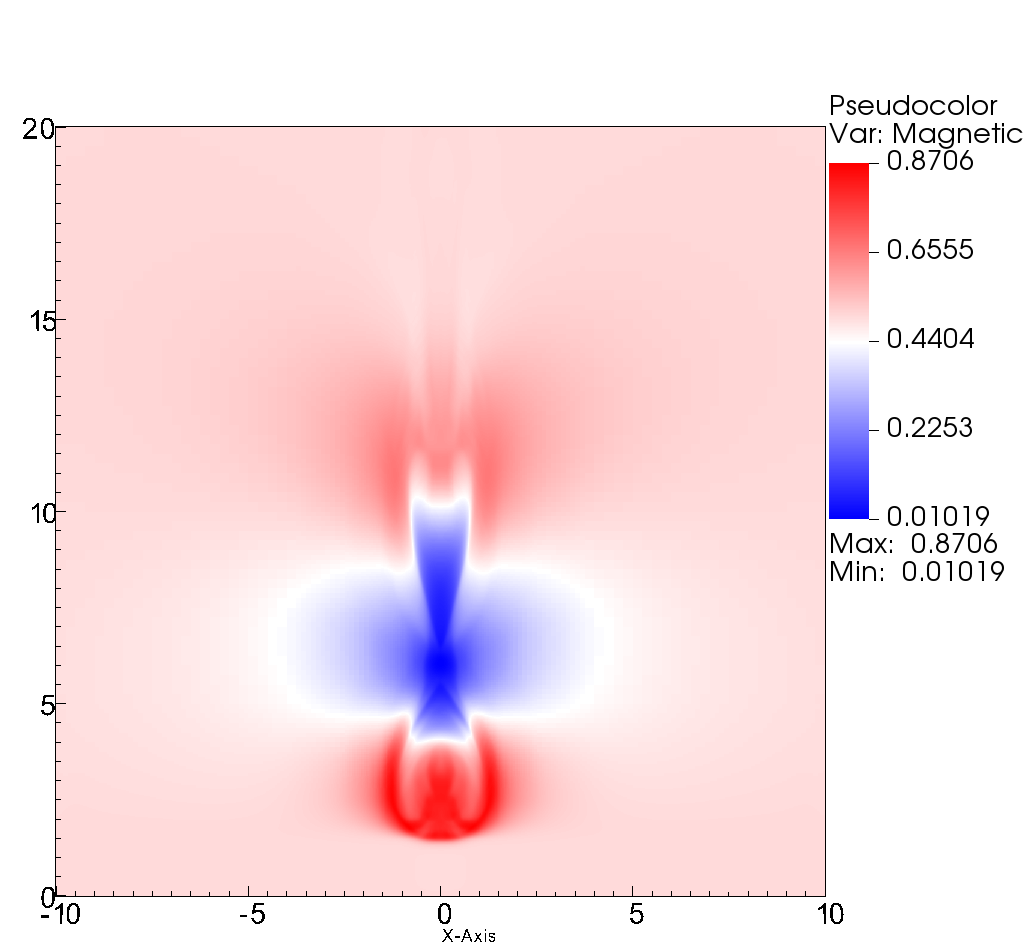}\\
\includegraphics[width=7.cm]{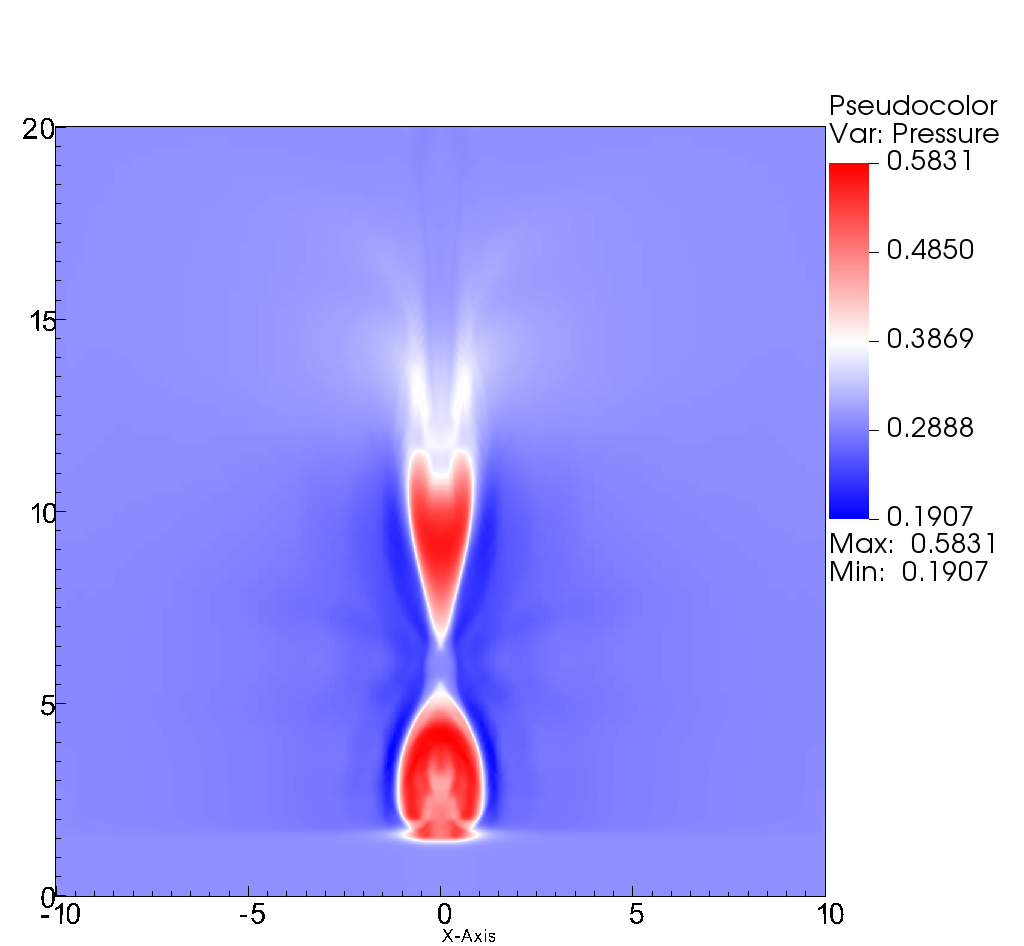}
\caption{\label{fig:Flare Pressure} 
Snapshot of the magnetic pressure (top) and the hydrodynamic pressure (bottom) of the LRCS test at $t=15$. }
\end{figure}

\begin{figure}
\centering
\includegraphics[width=7.cm]{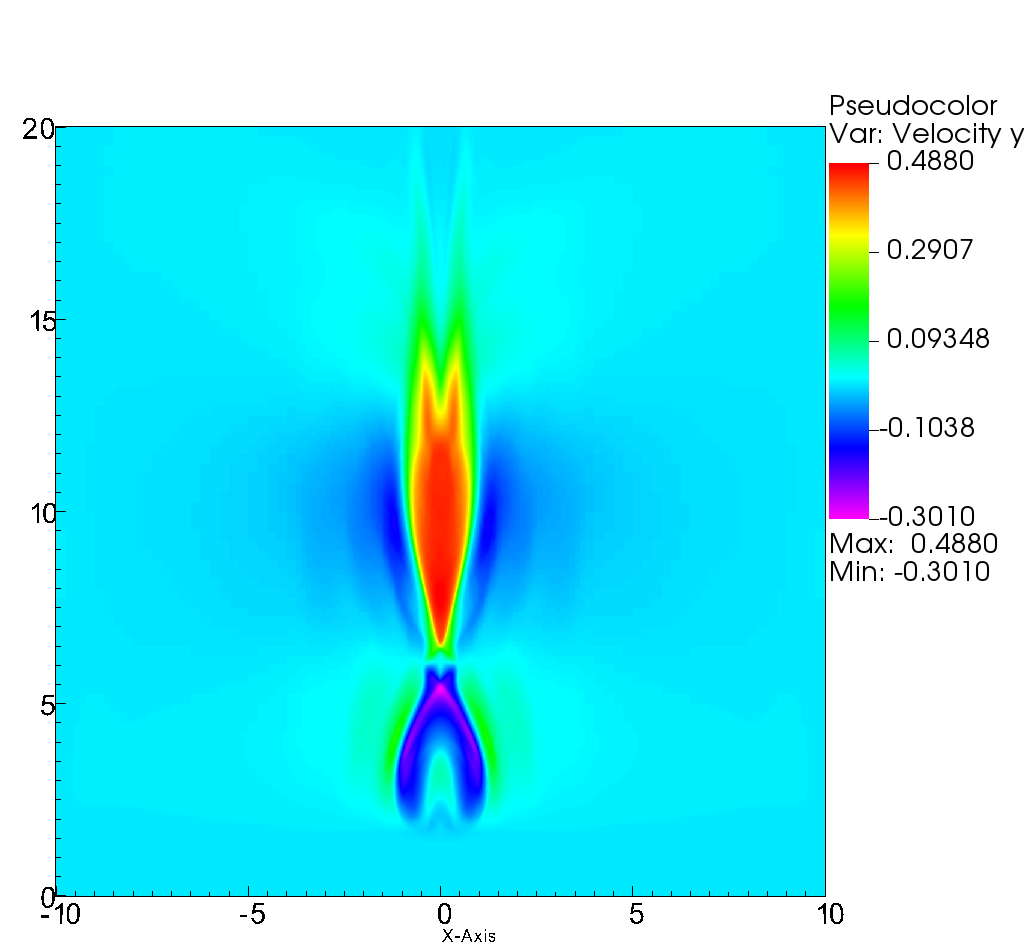}\\
\includegraphics[width=7.cm]{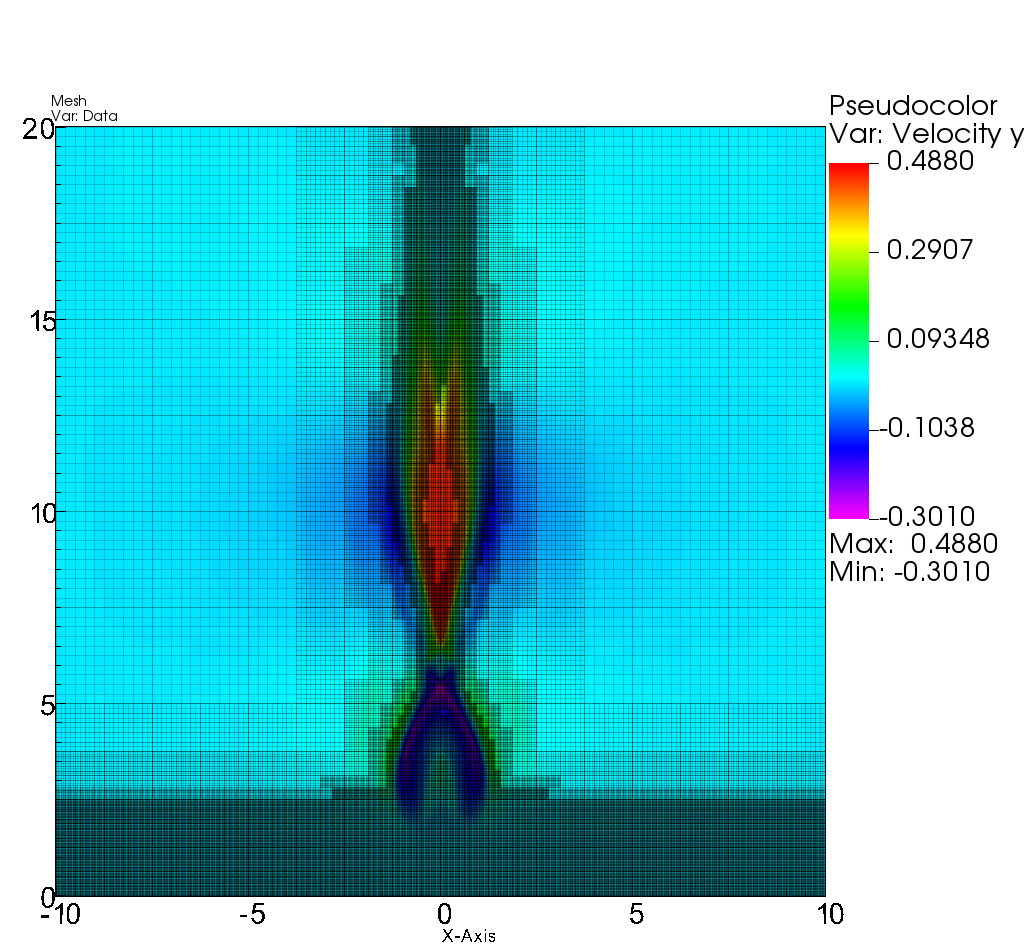}
\caption{\label{fig:Flare Vy lvl} 
Snapshot of $v_y$ (top) and the refinement structure (bottom) of the LRCS test at $t=15$.}
\end{figure}

In Figure \ref{fig:Flare Pressure} it is shown that near the maximum of the resistivity there is an increase in magnetic pressure as well as a decrease in hydrodynamic pressure. The results is that matter enters to this region from the sides and is expelled along the vertical directions as illustrated with the $y$ component of the velocity in Figure \ref{fig:Flare Vy lvl}. In this case we use precisely the velocity as the function to be tracked by the refinement criterion, which as seen in the plots, follows the discontinuities of $v_y$.

\subsection{Hydrodynamic Solar Wind}

Within the subject of Space Weather, the most commonly studied system is the Solar Wind, including the stationary wind.

In order to simulate this type of processes in the Solar System, we first solve the MHD equations with a null magnetic field in a cubic domain $[-x_{max},x_{max}]\times[-y_{max},y_{max}]\times[-z_{max},z_{max}]$, with $x_{max}=y_{max}=z_{max}$, centered at the position of the Sun, and consider $x_{max}$ bigger than the major semi-axis of Earth's orbit. For the position of Earth we use Heliocentric Earth Ecliptic (HEE) coordinates, in which Earth's position moves along the positive part of the $x-$axis from perihelion to aphelion.

Solar Wind simulations are carried out in a particular domain of interest, the volume contained from a sphere of $r_{in}\sim 20 R_{\odot}$ (see e.g. \citep{RileyGosling1998,GonzalezEsparza2003,Feng_2021}), until the faces of the cubic numerical domain. The reason to use a sphere of radius $r_{in}\sim 20R_{\odot}$ as an inner boundary is that at that distance from the Sun's surface, the characteristic speeds of the wind are all pointing outwards, and consequently the flow is in the wind regime, as opposed to the accretion regime which is possible at shorter distances. This allows the injection of the physical properties of a wind or a storm through this inner boundary without the danger of material entering back in an accretion process.

The permanent injection of plasma at the inner boundary can be simulated by maintaining the physical variables of the Solar Wind $\rho_{in},{\bf v}_{in},p_{in},{\bf B}_{in}$, with suitable values at $r=r_{in}$, constant in time. Notice however that in Cartesian coordinates this inner boundary is the boundary of a lego-sphere, not an actual sphere, and therefore the implementation of boundary conditions at each face of boundary cubes is  delicate  \citep{Kleinmann2009}. What we do to overcome this problem is to fill-in the whole lego-sphere with values of the primitive variables characterizing the wind at all times with the values at $r=r_{in}$, then the plasma will propagate outwards and eventually will fill the interplanetary space until it settles down to the heliosphere density and velocity profile \citep{Guzman_2022}.

At the external boundary, namely the faces of the cubic domain, we would like the fluid to continue its way out through the solar system and then use outflow boundary conditions at the outer cubic boundary.

As a particular example we simulate the formation of one of the stationary Solar Winds cases in \citep{GonzalezEsparza2003}, which is spherically symmetric, with no magnetic field, in a domain with inner boundary $r_{in}=17.18R_{\odot}$, density $\rho_{in}$=$2100~\mathrm{cm}^{-3}$, velocity ${\bf v}=v^r\hat{r}$, $v^r_{in}=2.5\times 10^5$m/s, temperature $T_{in}=5\times 10^5$K and zero magnetic field and adiabatic index $1.4$. We use the numerical domain to be $x_{max}=250R_{\odot}$ for the wind to expand and cover at least a volume that contains Earth's orbit.

{\it Translation from code to physical units}. Following \citep{Guzman_2022}, code and physical units are defined in terms of fixed scales of the state variables, for density $\rho_{phys}=\rho_0 \rho$, pressure $p_{phys}=p_0 p$, length $l_{phys}=l_0 l$, time $t_{phys}=t_0 t$, velocity $v_{phys}=v_0 v$, temperature $T_{phys}=T_0 T$. Notice that there is no restriction on density scale $\rho_0$, that we choose to be one.

We define code units by fixing the length scale to one solar radius
$l_0 = R_{\odot} = 6.9634\times 10^8 ~{\rm m}$, time to one hour
$t_0 = 3600~{\rm s}$, which defines the velocity scale
$v_0 = \frac{l_0}{t_0} = \frac{6.9634\times 10^8}{3600}~{\rm m/s}=1.93428\times 10^5 {\rm m/s}$. Density scale is fixed to $\rho_0 = \frac{m_H}{cm^3}=1.003573\times 10^{-21}{\rm kg/m^3}$ which defines the magnetic field scale 
$B_0 = v_0 \sqrt{\mu_0 \rho_0}$. Temperature scale is $T_0=m_{H} v_0^2 /k_B$ with $m_H=\mu m_P=0.6 \cdot 1.6726219\times 10^{27}$kg the Hydrogen molecular mass and $k_B$ Boltzmann's constant, and pressure and energy scales are $p_0=\rho_0 v_0^2$ and $E_0=\rho_0 v_0^2$. 

Initial conditions are available given for the density, velocity and temperature of the wind as follows:

\begin{eqnarray}
\rho_0({\bf x}) &=& \left\{ \begin{array}{ll}
					\rho_{sw}, &~~~ {\rm if}~r<r_{in}\\
					&\\
					0.01\rho_{sw}, &~~~ {\rm if}~r>r_{in}
					\end{array} \right. \nonumber\\
{\bf v}_0({\bf x}) &=&  \left\{ \begin{array}{ll}
					(x v_{sw}/r, y v_{sw}/r, z v_{sw}/r), & ~~~{\rm if}~r<r_{in}\\
					&\\
					0, & ~~~{\rm if}~r>r_{in}
					\end{array} \right.\nonumber\\
					&&\nonumber\\
T_0({\bf x}) &=& T_{sw} \label{eq:ics}
\end{eqnarray}

\noindent where $r=\sqrt{x^2+y^2+z^2}$. The pressure needed as the primitive variable instead of temperature, is calculated from $p_0({\bf x})=\rho_0({\bf x})  T_0({\bf x}) $, then internal energy is given by the equation of state $e({\bf x}) =p_0({\bf x})/\rho_0({\bf x}) /(\gamma - 1)$, and finally the total energy is $E=\rho(\frac{1}{2}v^2+e)$.

The simulation uses a base resolution of $80\times80\times 80$ cells with 3 refinement levels, producing an equivalent resolution of $640\times640\times 640$ cells. We use the refinement criterion \eqref{RcritMio} with a threshold value  $\chi_r=10$. Outflow boundary conditions are set at the borders of the numerical domain. Numerical fluxes are computed using the HLLE formula along with the  mimod reconstructor. A CFL factor of $0.125$ was used throughout the whole simulation.

\begin{figure}
\centering
\includegraphics[width=5cm]{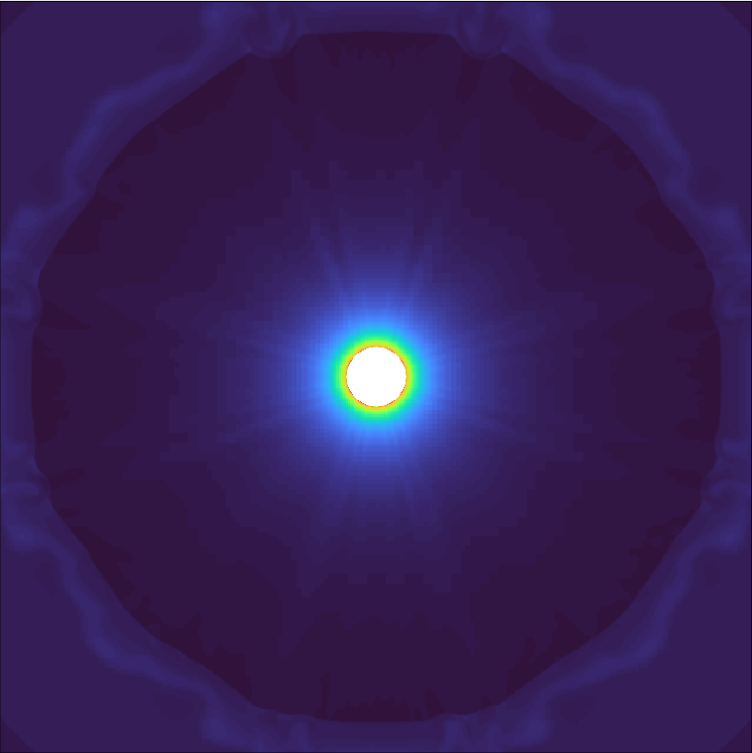}\\
\includegraphics[width=5cm]{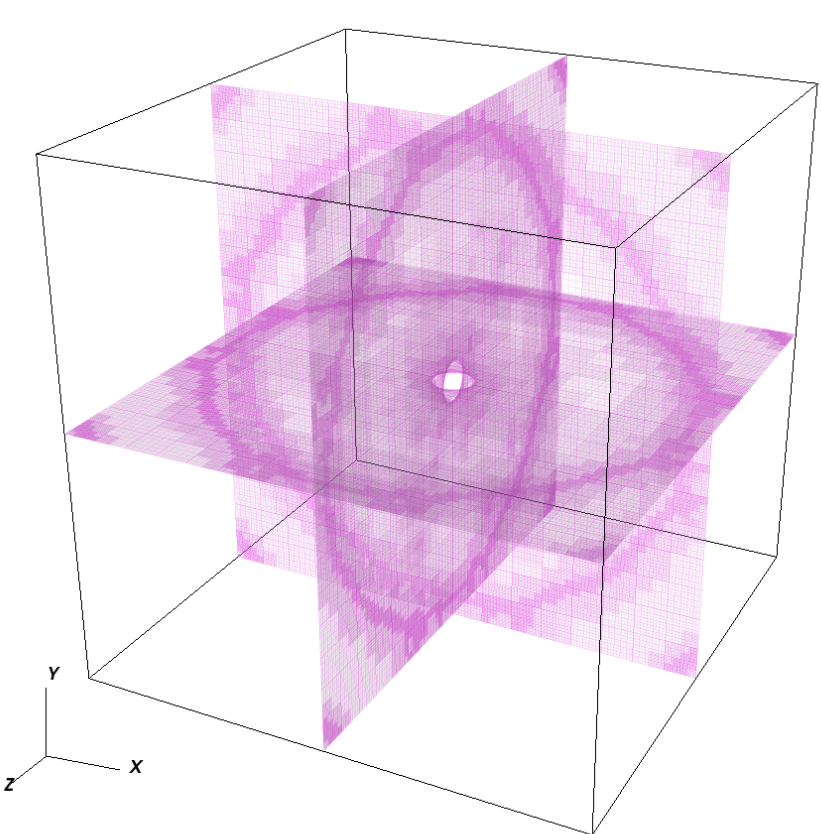}
\caption{ Snapshots of density and its AMR tracking at $t=234$hr of the hydrodynamic solar wind test, before the wind achieves a stationary state. The initial shock wave as well as the contact discontinuities are covered by finer grids.}\label{fig:ssw}
\end{figure}

At initial time the wind variables have constant values within the sphere of radius $r_{in}$, which simulates a stationary pumping of plasma into the heliosphere. The gas fills the numerical domain with the front wave that propagates outwards until it reaches the external boundary and leaves the domain. In Figure \ref{fig:ssw} we show how the refinement criteria follows this front wave. Once the wave leaves the boundary the flow approaches a stationary state.

\begin{figure}
\centering
\includegraphics[width=6cm]{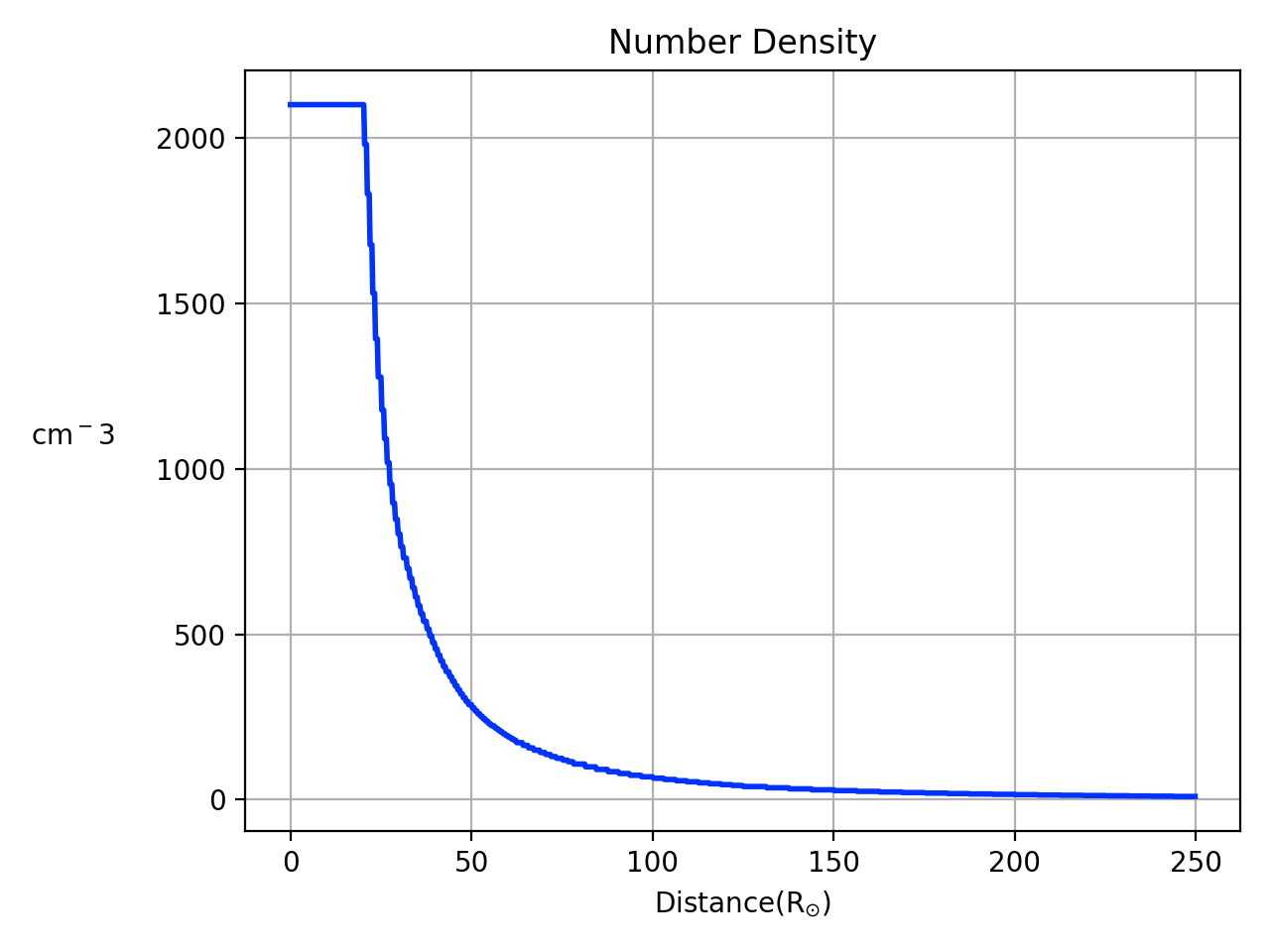}\\
\includegraphics[width=6cm]{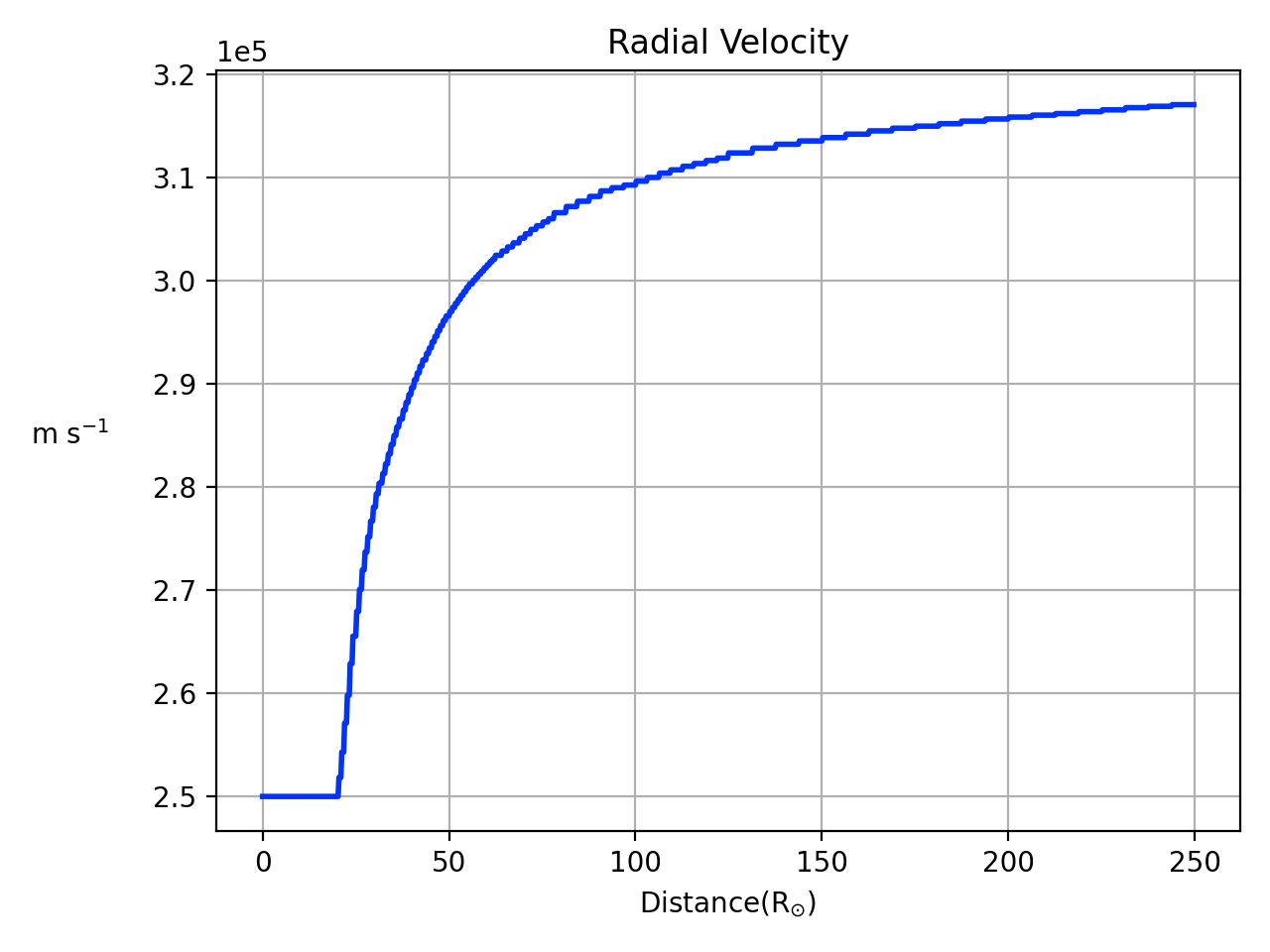}\\
\includegraphics[width=6cm]{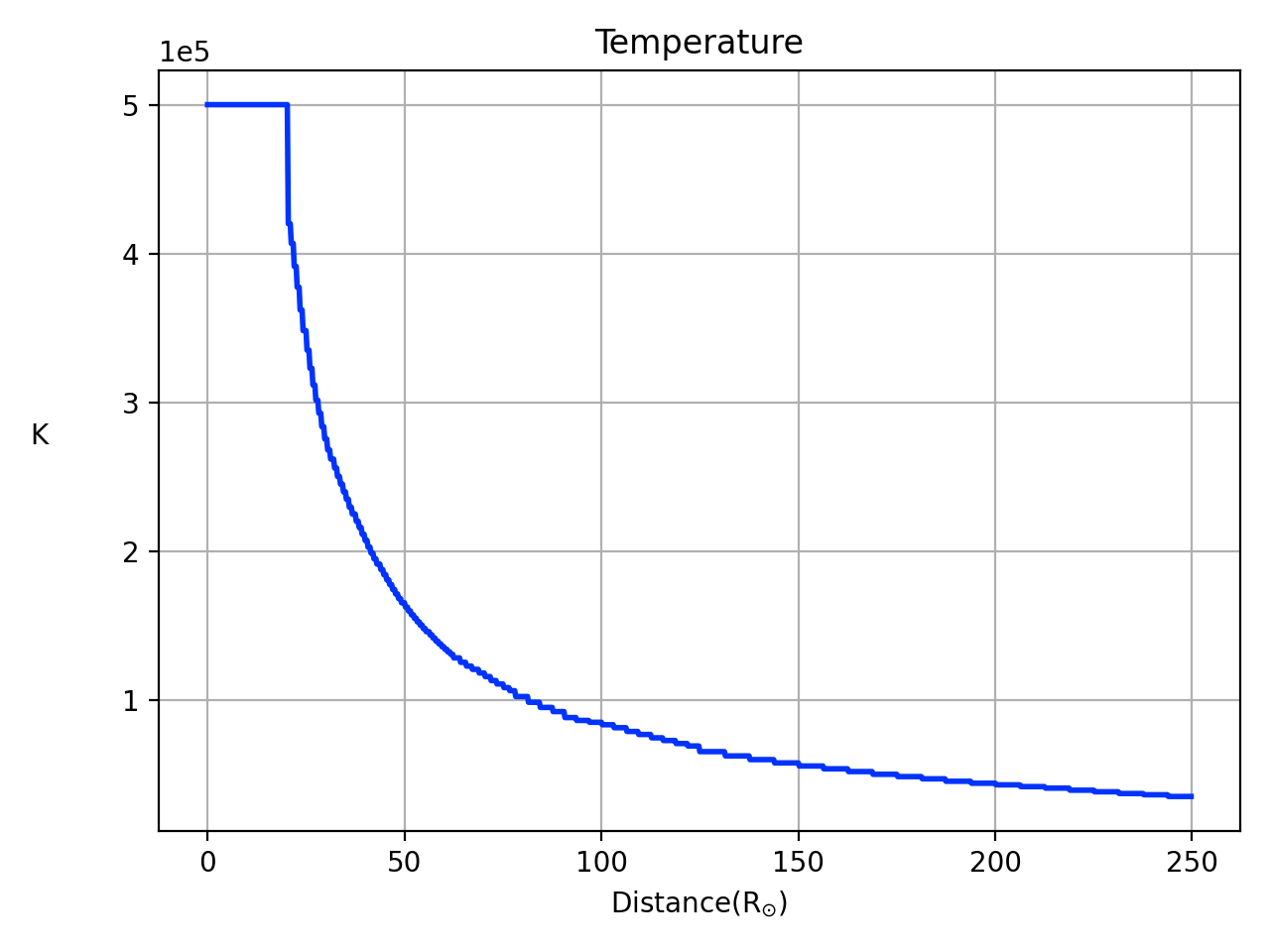}
\caption{Density, radial velocity and Temperature of the Hydrodynamic Solar Wind along the $x-$axis after the flow has become stationary.}\label{fig:QSSW_DTVr}
\end{figure}

The result of this simulation appears in Figure \ref{fig:QSSW_DTVr}, where we show snapshots of density, $x$ component of ${\bf v}$ and temperature of the gas along the $x-axis$, until the flow stabilizes in the whole domain by $t\sim 1000$ hours. In order to have a reference of the spatial domain, in these coordinates (HEE) the average Earth location is $x_{Earth}=214.83452R_{\odot}\hat{x}$.

\begin{figure}
\centering
\includegraphics[width=8.cm]{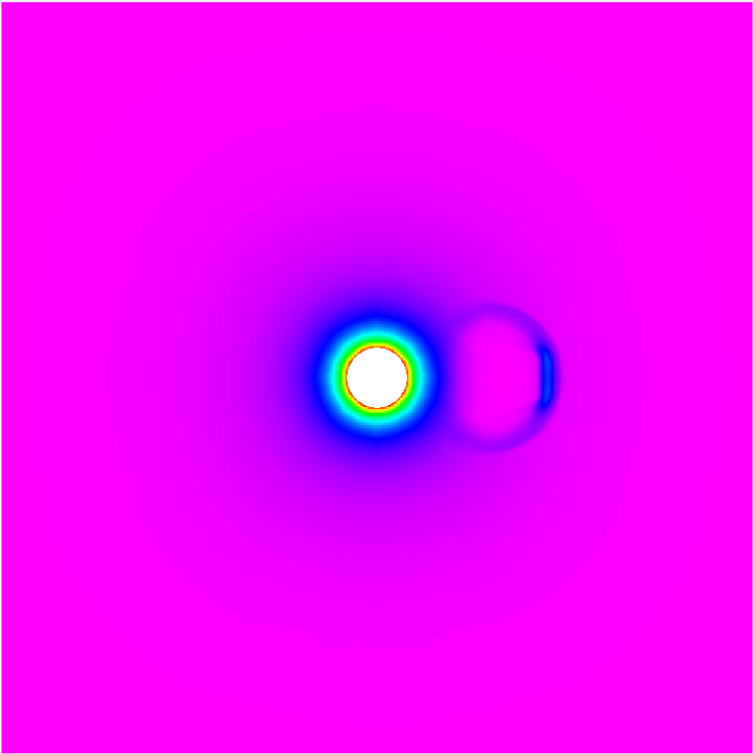}\\
\includegraphics[width=2.6cm]{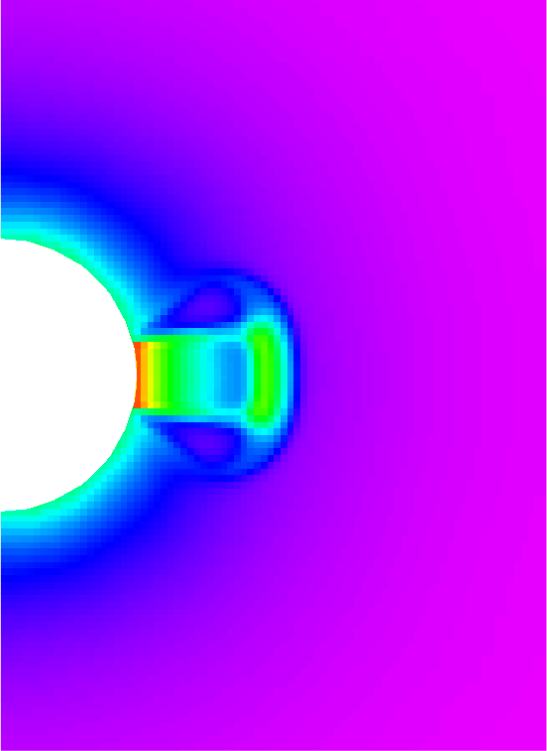}
\includegraphics[width=2.6cm]{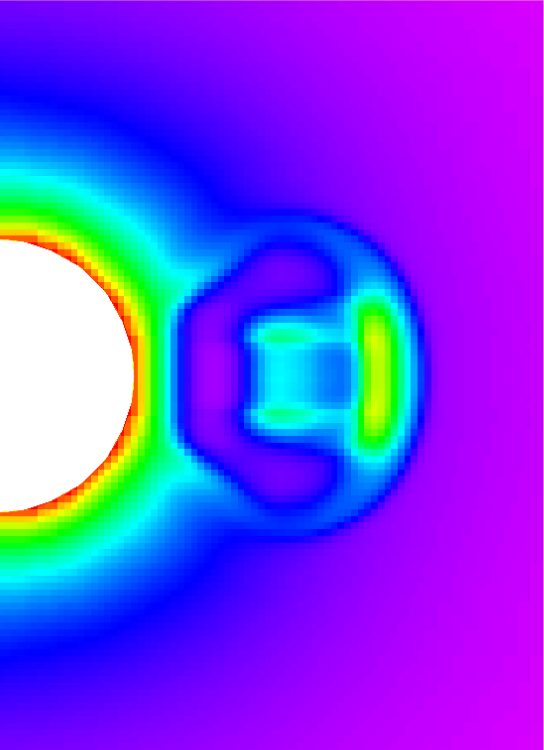}
\includegraphics[width=2.6cm]{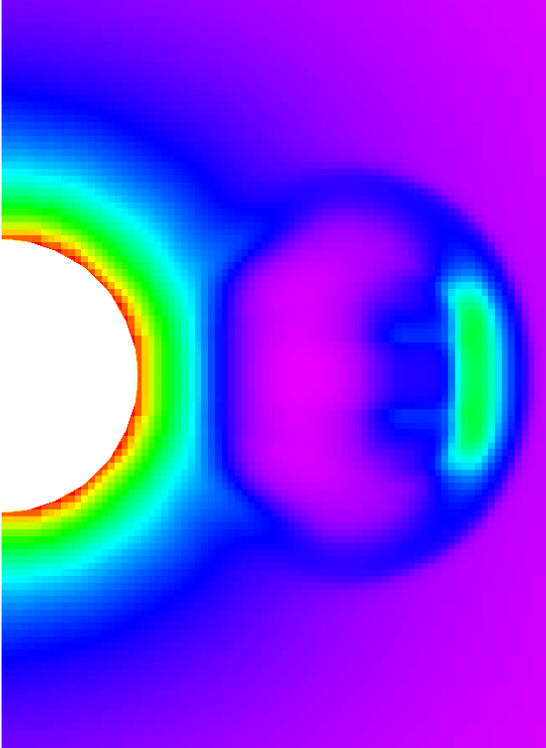}
\caption{ Snapshots of the density profile on the $xy-$plane after the perturbation has been injected into the stationary solar wind configuration at time $t=34,8$hr (top), $t=7.50$hr, $13.98$hr, $19.01$hr (bottom).}\label{fig:sswCME}
\end{figure} 

\begin{figure}
\centering
\includegraphics[width=7cm]{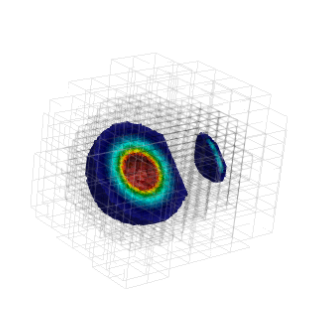}
\caption{ Refinement structure after the internal boundary perturbation has been injected into the stationary solar wind configuration at $t=33,8$hr. In this figure we also overlap a threshold selection of density corresponding to the internal boundary as well as the CME perturbation.}\label{fig:CME lvl}
\end{figure} 

After the wind has become stationary, we inject a toy-model of  Coronal Mass Ejection (CME). We characterize the CME launched from the inner boundary at $r = 20R_\odot$ by its density, velocity and widening angle $\theta=\pi/6$.
The injection is implemented similarly as the stationary Solar Wind at the inner boundary provided density, velocity and temperature of the CME. As a test we inject a toy CME with the following properties $\rho_{\mathrm{CME}} = 3\times \rho_\mathrm{{wind}}$ 
$v_{r,\mathrm{CME}} = 2.5 \times v_{r,\mathrm{wind}}$ and $T_{\rm CME}=2\times T_{\mathrm{wind}}$, conditions taken from  \citep{GonzalezEsparza2003}.
We launch this CME centered at the solar equator, during a time window of 3 hours. In Figure \ref{fig:sswCME} we show snapshots of density during the propagation of the CME. With this initial setup the initial burst propagates, its density profile smears out and  widens. In Figure \ref{fig:CME lvl}  we show that the structure of the highest refinement level tracks the perturbation.


\subsection{Tilted Magnetic Dipole Solar Wind}

Beyond the solar corona, the solar wind is fairly well and commonly modelled using ideal MHD. In this context, the macroscopic plasma quantities, such as density, velocity and temperature, can all be related to the topology of the field lines of the magnetic field. Current techniques applied to simulate the interplanetary heliosphere rely on this fact, for example the WSA \citep{arge2004stream} approximation and its implementation in the ENLIL model \citep{odstrcil2003modeling}.

In order to explain our implementation of the magnetic field we briefly review the bases of empirical models. The physical domain is partitioned into some basic  spherical shells approximately located at given spheres: the photosphere at 1 $\mathrm{R}_{\odot}$, the Solar Corona (SC) at  $\mathrm{R}_{SC}\approx 2.5 \mathrm{R}_\odot$, the inner boundary at $R_{in}\approx20\mathrm{R}_\odot$, and the rest of the numerical domain that includes a volume containing at least the Earth's orbit, of $\approx 1 \mathrm{AU}$. The magnetic field at the photosphere is reconstructed from magnetogram synoptic maps, used as input to construct analytic expressions of the magnetic field at the Corona via the potential field source surface method \citep{newkirk1969magnetic}. Next, the \textit{super-radial expansion factor} ($f_s$), that refers to the variation of field strength along a magnetic flux tube is calculated, for example as done in \citep{shiota2014inner}:

\begin{equation}\label{sr expansion factor}
f_s(\theta_{sc},\phi_{sc})=\left(\frac{R_\odot}{R_sc}\right)^{2}\frac{B_r(R_\odot,\theta_{\odot},\phi_{\odot})}{Br(R_{sc},\theta_{sc},\phi_{sc})},
\end{equation}

\noindent where $B_r$ is the radial component of the magnetic field. The coordinates $(R_{\odot},\theta_\odot,\phi_\odot)$ of the foot point of the magnetic flux tube are located by using a first order integration of the magnetic field lines taking $(R_{sc},\theta_{sc},\phi_{sc})$ as the starting point; if, after a finite number of steps, the foot point is not found, a $f_s$ default value is chosen so that the radial velocity is minimal according to \eqref{WS velocity}.

Once $f_s$ is calculated, we use the empirical  Wang and Sheely formula  \citep{wang1990solar, arge2000improvement} that gives the wind radial velocity at the inner boundary as

\begin{equation}\label{WS velocity}
    v_r(R_{in},\theta_{in},\phi_{in})=267.5+\frac{410}{f_s^{0.4}(\theta_{in},\phi_{in})}\mathrm{km/s}.
\end{equation}

\noindent With this in mind, we have woven a test based on the empirical construction of the solar wind at the inner boundary using an analytical expression of the magnetic field.

Based on the Parker Solar Wind model \citep{priest2014magnetohydrodynamics}, we use a constant temperature, as well as a magnetic field that fulfills conservation across radial flux tubes:

\begin{equation}\label{Inner Boundary Magnetic Field}
    B_r(R_{in})=\left(\frac{R_{sc}}{R_{in}}\right)^2 B_r(R_{sc}).
\end{equation}

To obtain the solar wind radial velocity profile at the inner boundary in equation \eqref{WS velocity}, we impose the magnetic field of a tilted dipole given by

\begin{equation}\label{Tilted Magnetic field}
    \mathbf{B}(\mathbf{r})=\frac{3\mathbf{r}(\mathbf{m}\cdot\mathbf{r})}{r^5}-\frac{\mathbf{m}}{r^3},
\end{equation}

\noindent where $\mathbf{m}$ is the magnetic dipole moment. We use an inertial heliographic reference frame \citep{burlaga1995interplanetary} that does not have Coriolis forces due to rotation. In this sense, the internal boundary expresses an ejected hot sphere of magnetized plasma that rotates like a solid. To apply the rotation at the internal boundary, we transform the $m_{\phi}$ component of the magnetic dipole by $\phi\rightarrow{\phi+\Omega_{\odot}t}$, where $\Omega_{\odot}$ is the angular frequency of the Sun. This will rotate the inner magnetic field on equation \eqref{Tilted Magnetic field} and, because all of the boundary quantities are connected to the magnetic field, they will also experience the same rotation.

The number density $n$ is given by a measure of the radial flux of number density at 1 AU in slow wind conditions \citep{smith2009heliophysics},

\begin{equation}\label{Inner Boundary number density}
    n(R_{in})=n_0\left(\frac{R_{0}}{R_{in}}\right)^2\times \frac{v_0}{v_r(R_{in})},
\end{equation} 

\noindent where $n_0$ $8.06\times 10^6 \mathrm{cm}^{-3}$, $V_0=267.5\mathrm{km/s}$, $R_0=1\mathrm{AU}$. Finally the hydrodynamic pressure can be obtained from the equation of state of the ideal gas.

The simulation specifications for this test are the following: The computational domain is the cubic box of size $256R_{\odot}\times256R_{\odot}\times256 R_{\odot}$, centered at the Sun's location. The initial resolution is given by $128^3$ cells with three refinement levels, which corresponds to an equivalent resolution of $1024^3$ cells. We use outflow boundary conditions at the edges of the numerical domain and set the internal boundary at $R_{in}=20R_{\odot}$. Time stepping is handled with second order RK with a CFL factor of  $0.125$; numerical fluxes are obtained through the HLLE approximation. The initial temperature of the system is set to $1\times10^{5}K$, with an adiabatic index of $\gamma=5/3$. The magnetic dipole strength is $m=1\times10^{-5}$T $\mathrm{m}^{-3}$, its colatitud inclination angle is $\theta=\pi/5$ and the magnetic field at the inner Boundary is given by the radial component of \eqref{Tilted Magnetic field}.

We use two types of refinement, a fixed one and an adaptive one. In order to minimize the effect of geometrical errors due to the inner boundary to be a lego sphere, we set a fixed mesh refinement near the boundary so that in this region we use maximum spatial and temporal resolution. The adaptive refinement is a mixture of the one proposed by \citep{matsumoto2019dynamical}, where we refine a data block if there is a change in polarity in $B_r$, and, because we desire to capture the spiral-like behavior in the radial component of the velocity, we choose $\sigma(\mathbf{U})=v_r$ at equation \eqref{ref criteria} with a threshold $\chi_r <0.1$. This refinements are implemented only in regions where $|z|<2 R_{in}$ so that only the heliospheric current sheet is refined.

\begin{table}
    \centering 
    \begin{tabular}{cc}
    \hline
    Quantity & Normalization \\
    \hline
    Time & $t_0=3600$ s\\
    Length & $L_0=R_{\odot}=6.9634\times 10^{8}$ m\\
    Velocity & $v_0=L_0/t_0\approx 1.93\times 10^5$m/s\\
    Temperature & $m_H v_0^2/k_B\approx2.73\times10^6$K\\
    Number Density & $N_0=1\times10^9\mathrm{m}^{-3}$ \\
    Density & $\rho_0=m_H N_0\approx1.004\times10^{-18}$ \\
    Magnetic Field & $B_0=v_0 \sqrt{\mu_0 \rho_0}\approx2.1786\times10^{-7}$ Tesla\\
    \hline
    \end{tabular}     
    \caption{Normalization of constants for the tilted magnetic dipole test are  $m_H=0.6\times 1.6726219\times10^{-27}$kg the mean hydrogen molecule mass, $k_{B}=1.3806488\times10^{-23}$J/K the Boltzmann constant, and $\mu_0=1.256637\times10^{-6}\mathrm{N/A}^2$ the magnetic permeability.}
    \label{tab:Normalization Tilted-Dipole}
\end{table}

\begin{figure}
    \centering
    \includegraphics[scale=0.24]{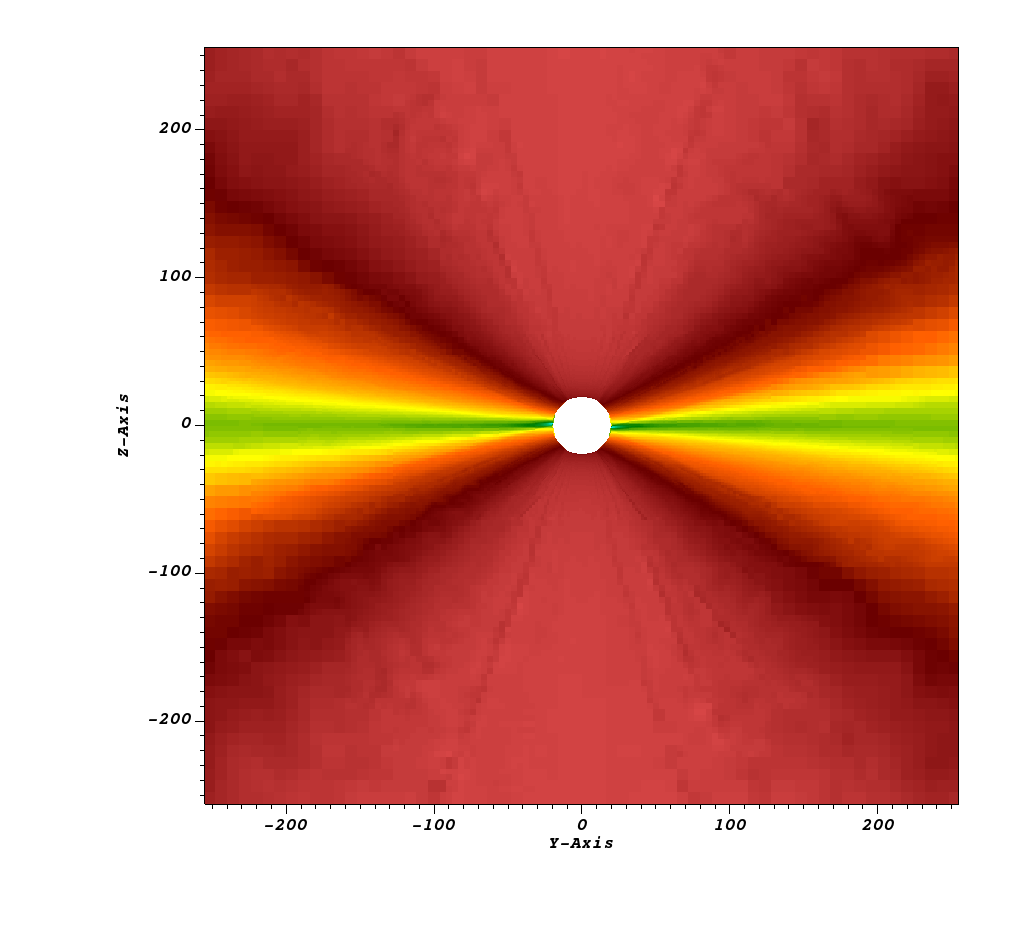}
    \caption{Plane cut at $x=0$ of the radial velocity distribution at time $t=200$h of the tilted magnetic dipole solar wind test. Velocity ranges  are from $\approx 2.8\times10^5$m s$^{-1}$ (green) to $\approx 5.56\times10^5$m s$^{-1}$ (dark red).}
    \label{fig:Tilted_Dipole_RadialV_x0}%
\end{figure}

\begin{figure}
    \centering
    \includegraphics[scale=0.24]{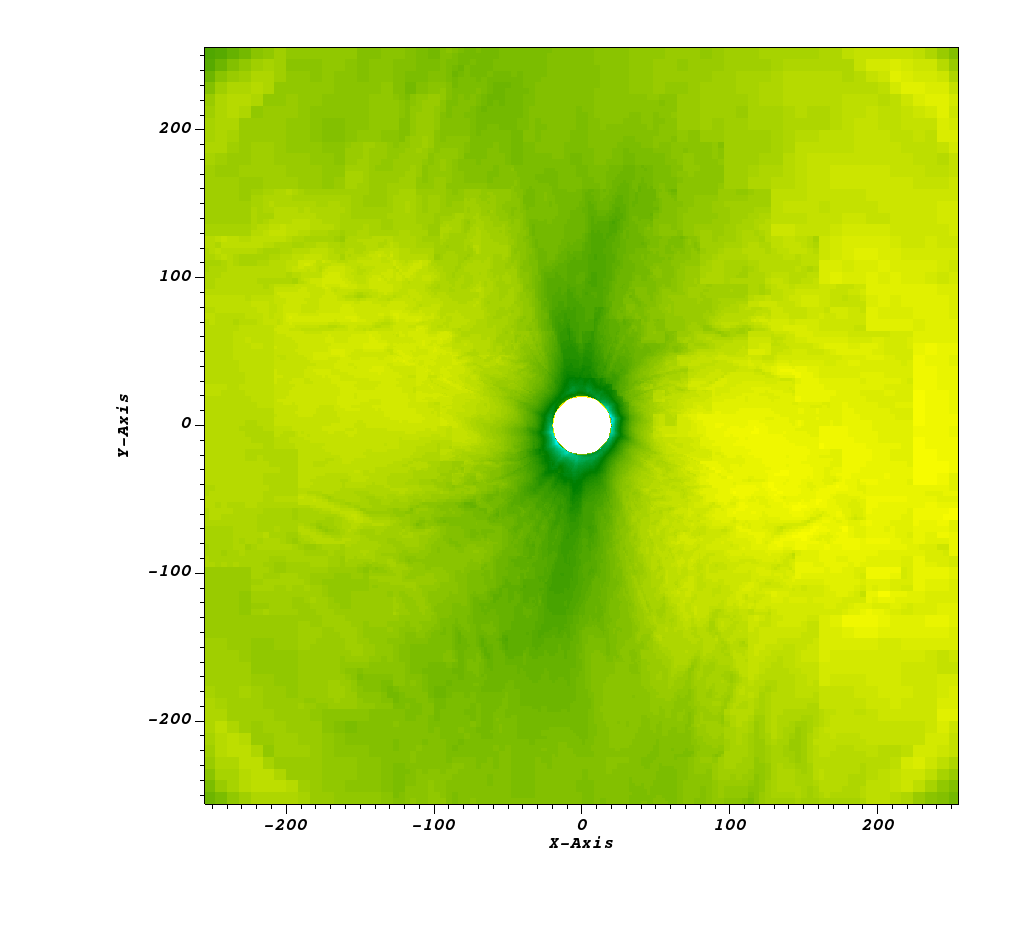}
    \caption{Projection of the radial velocity distribution at the plane $z=0$ and time $t=200$h of the tilted magnetic dipole solar wind test. Velocity ranges  are from $\approx 2.8\times10^5$m s$^{-1}$ (dark green) to $\approx 3.25\times10^5$m s$^{-1}$ (light green).}
    \label{fig:Tilted_Dipole_VelR_z0}
\end{figure}

In order to simulate this scenario, we connect physical and code units according to the values in Table \ref{tab:Normalization Tilted-Dipole}. In Figure \ref{fig:Tilted_Dipole_RadialV_x0} we show a meridian cut of the radial velocity, and notice that there are two regions corresponding to a slow streamer belt and a fast wind region near the poles. Figure \ref{fig:Tilted_Dipole_VelR_z0} shows the rotation effect at the zenith cut of the velocity, which develops a gradient in its magnitude typically associated with co-rotating interaction regions.

\begin{figure}
    \centering
    \includegraphics[scale=0.3175]{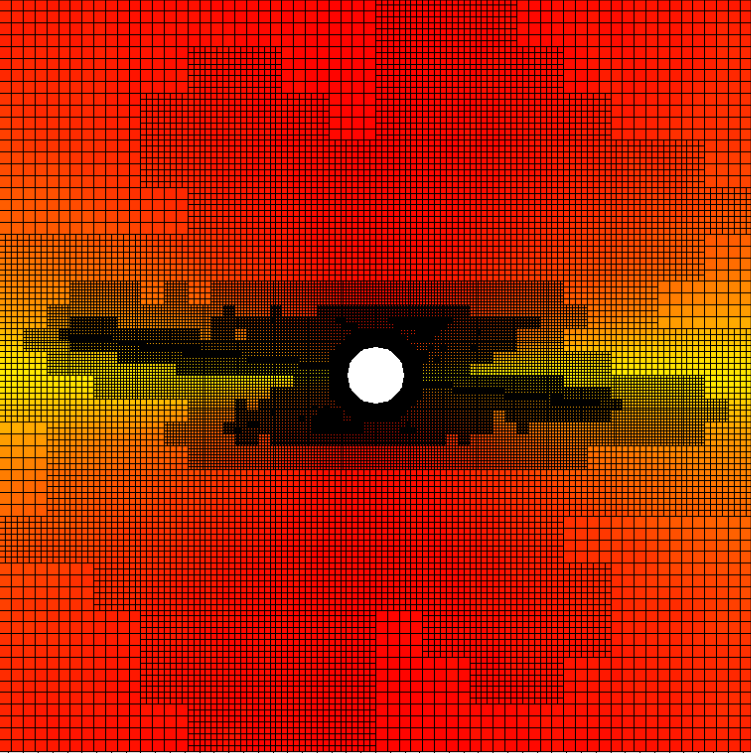}
    \caption{Plane cut at $x=0$ of the mesh refinement on the radial velocity distribution at $t=200$h of the tilted magnetic dipole solar wind test. The z-shaped refined zone is due to the polarity change of the radial component of the magnetic field. The vertical fringes refined outside of $|z|<2R_{in}$ are due to the criterion based on radial velocity.}
    \label{fig:Tilted_Dipol_VelR_mesh}
\end{figure}

In reference to mesh refinement, Figure \ref{fig:Tilted_Dipol_VelR_mesh} shows the various domains and the resolutions used in each patch, 
at the plane $x=0$. The refinement tracks the slower wind at low latitudes as  expected because it is also the region where there is a shift on polarity in the magnetic field.

\begin{figure}
    \centering
    \includegraphics[width=8.cm]{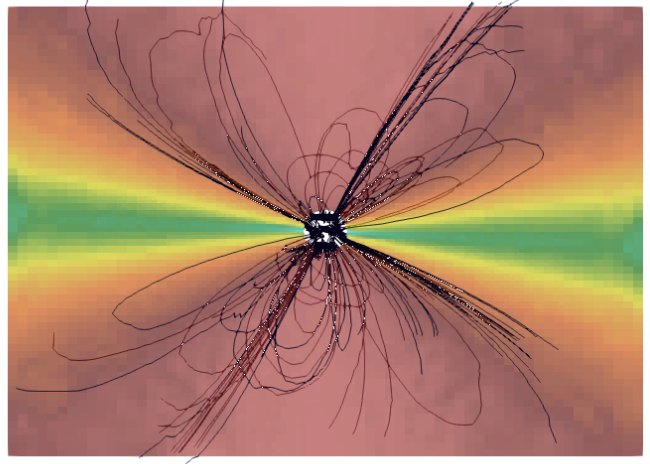}
    \caption{Magnetic field lines of the tilted magnetic dipole solar wind test at $t=200$, overlapped with the plane cut of constant azimuth angle $\Phi=3\pi/2$ of the radial velocity distribution.Velocity ranges  are from $\approx 2.8\times10^5$m s$^{-1}$ (green) to $\approx 5.56\times10^5$m s$^{-1}$ (dark red).} 
    \label{fig:Tilted_Dipole_Fieldlines}
\end{figure}

At big interplanetary distances, the magnetic profile of the solar wind is dominated by the dipole moment. Figure  \ref{fig:Tilted_Dipole_Fieldlines} shows the magnetic field lines over the whole domain; it can be seen that near the equator, field lines are sporadic, as expected for a magnetic dipole. Certain torsion of the field lines associated with the rotation of the injected boundary can also be noticed. This simulation achieves this specification and also manages to evolve the rotating magnetized solar wind that generates the Parker spirals. This is a test case toward the real more elaborate simulations in practical forecasting, which involve specialized PFSS definitions to extend the range of heliosphere simulations even further.

The issue we are tackling on this test, in the context of solar wind simulations, is of augmenting spatial resolution not only at the internal boundary but also capturing dynamical structures at regions of interest farther away. Certainly AMR on rectilinear grids is not the only strategy. For example, in the project of sun to earth simulations \citep{narechania2021integrated}, AMR regriding strategies are employed using hexahedral blocks, obtaining high resolution on the formation of current sheets as well as on the propagation of CME, at the cost of needing special care of the numerical fluxes at intercell boundary faces. Another case is the ICARUS code \citep{verbeke2022icarus}, which achieves high  resolution simulations using AMR on spherical grids, which do not have as complex flux formulas as the hexahedral simulations, but due to the geometry of the equations there is a need of special care of singularities of the equations at the poles. AMR on rectilinear grids do not have neither of this problems, but the issue of having a lego sphere at the internal boundary is a hard problem, it overburdens the usage of the PFSS schemes and the use of co-rotating frames might also induce non physical phenomena; major improvements on this test can be made, in this sense, by applying corrections which could mitigate this geometric problem, for example with the Immersed Boundary method \citep{mittal2005immersed} or averaging at the excision surface \citep{Kleinmann2009}, as in the hydrodynamic solar wind tests from the previous section.

\section{Final comments and conclusions}
\label{sec:conclusions}

In this paper we have presented a new implementation that uses AMR to solve the MHD equations, with the expectation to have a theoretical laboratory to study and contribute to Space Physics at heliosphere scales.

CAFE-AMR uses simple and standard numerical methods that allow the simulation of various processes of importance, which are concentrated in a series of idealized standard test problems. These tests show the ability of CAFE-AMR to track well known features prominent in solar physics, with particular emphasis on the adaptability of refined domain patches:
Standard ideal MHD tests of the Kelvin-Helmholtz and the Rayleigh-Taylor instabilities that are ubiquitous in solar physics and generate highly localized phenomena; the Localized Resistivity current sheet which simulates the evolution of solar flares in regions near the chromosphere-corona interface; heliospheric related tests include the formation and stabilization of a  non-magnetized Solar Wind and CME propagation as well as the formation and stabilization of a magnetized Solar Wind which generates a rotating current sheet.

Further work will involve specific applications that include convection and radiation with the scope of modeling chromospheric events and non-ideal MHD effects. The code in this paper is expected to be enriched with the methods and scenarios involving thermal conduction \cite{CAFEQ,SpiculesQ} and radiation \citep{CAFER}.


\section*{Acknowledgements}

Ricardo Ochoa Armenta receives support from CONACyT under CVU 785828. This research is supported by grant CIC-UMSNH-4.9 of Universidad Michoacana. The runs were carried out in the Big Mamma cluster of the Laboratorio de Inteligencia Artificial y Superc\'omputo, IFM-UMSNH.

\section*{Data availability}

The data and code underlying this article will be shared on reasonable request to the corresponding author.\\


\bibliographystyle{mnras}

\bibliography{CodePaper} 



\bsp	
\label{lastpage}
\end{document}